\documentclass[twocolumn]{aastex631}

\usepackage{graphicx}
\usepackage{savesym}
\savesymbol{tablenum}
\usepackage{siunitx}
\restoresymbol{SIX}{tablenum}
\usepackage{amsmath}
\usepackage{natbib}

\usepackage{hyperref}
\usepackage{orcidlink}
\usepackage{subfigure}

\begin{document}

%% Reintroduced the \received and \accepted commands from AASTeX v5.2
%\received{XXX}
%\revised{YYY}
%\accepted{ZZZ}
%% Command to document which AAS Journal the manuscript was submitted to.
%% Adds "Submitted to " the argument.
% \submitjournal{ApJ}

%\submitjournal{AAS}

\author{Ye-Peng Yan}
\affiliation{Key Laboratory of Particle Astrophysics, Institute of High Energy Physics, Chinese Academy of Sciences, 19B Yuquan Road, Beijing 100049, People’s Republic of China}

\author{Si-Yu Li}
\affiliation{Key Laboratory of Particle Astrophysics, Institute of High Energy Physics, Chinese Academy of Sciences, 19B Yuquan Road, Beijing 100049, People’s Republic of China}

\author{Yang Liu}
\affiliation{Key Laboratory of Particle Astrophysics, Institute of High Energy Physics, Chinese Academy of Sciences, 19B Yuquan Road, Beijing 100049, People’s Republic of China}

\author{Jun-Qing Xia}
\affil{Institute for Frontiers in Astronomy and Astrophysics, Beijing Normal University, Beijing 100875, China}
\affil{School of Physics and Astronomy, Beijing Normal University, Beijing 100875, P.R.China}

\author{Hong Li}
\affiliation{Key Laboratory of Particle Astrophysics, Institute of High Energy Physics, Chinese Academy of Sciences, 19B Yuquan Road, Beijing 100049, People’s Republic of China}
\affiliation{University of Chinese Academy of Sciences, Beijing 100049, People's Republic of China}

\correspondingauthor{Si-Yu Li, Hong Li}
\email{lisy@ihep.ac.cn, hongli@ihep.ac.cn}

\shortauthors{Yan et al.}

%\correspondingauthor{Si-Yu Li}
%\email{ptanouri@phas.ubc.ca}

%\title{Foreground removal with Transformer model for AliCPT-1}
\title{Foreground Removal in Ground-Based CMB Observations Using a Transformer Model}
\begin{abstract}
%We present a deep learning method based on the Transformer model, named Transformer for Cosmic Microwave Background (\texttt{TCMB}), which is designed to effectively process HEALPix spherical maps. This method has been applied to foreground subtraction of polarization in ground-based CMB observations. Using simulated data, the \texttt{TCMB} method demonstrates robust performance in removing foreground contamination. The mean absolute variance for the reconstruction of the noisy CMB Q map is measured at $0.0081\pm 0.0036\ \mu$K, while for the noisy CMB U map, it is $0.0073\pm 0.0039\ \mu$K. To mitigate biases caused by instrumental noise, a cross-correlation approach using two half-mission maps was employed, successfully recovering CMB EE and BB power spectra that align closely with the true values. These findings validate the effectiveness of the \texttt{TCMB} method. Furthermore, we conducted a comparative analysis between the \texttt{TCMB} method and a convolutional neural network (CNN)-based approach. The results show that this new method outperforms the CNN method in both foreground removal and the processing of HEALPix spherical sky maps.

We present a novel method for Cosmic Microwave Background (CMB) foreground removal based on deep learning techniques. This method employs a Transformer model, referred to as \texttt{TCMB}, which is specifically designed to effectively process HEALPix-format spherical sky maps. \texttt{TCMB} represents an innovative application in CMB data analysis, as it is an image-based technique that has rarely been utilized in this field. Using simulated data with noise levels representative of current ground-based CMB polarization observations, the \texttt{TCMB} method demonstrates robust performance in removing foreground contamination. The mean absolute variance for the reconstruction of the noisy CMB Q/U map is significantly less than the CMB polarization signal. To mitigate biases caused by instrumental noise, a cross-correlation approach using two half-mission maps was employed, successfully recovering CMB EE and BB power spectra that align closely with the true values, and these results validate the effectiveness of the \texttt{TCMB} method. 
Compared to the previously employed convolutional neural network (CNN)-based approach, the \texttt{TCMB} method offers two significant advantages: (1) It demonstrates superior effectiveness in reconstructing CMB polarization maps, outperforming CNN-based methods. (2) It can directly process HEALPix spherical sky maps without requiring rectangular region division, a step necessary for CNN-based approaches that often introduces uncertainties such as boundary effects. This study highlights the potential of Transformer-based models as a powerful tool for CMB data analysis, offering a substantial improvement over traditional CNN-based techniques.

% \st{Some overstruck text}
\end{abstract}

%%%%%%%%

\section{Introduction} 
\label{sec:intro}

The Cosmic Microwave Background (CMB) provides crucial observational information for precision cosmology \citep[e.g.,][]{Bennett2003,Story2013,Das2014,PlanckCollaboration2016}. The next focus of research lies in the precise measurement of CMB polarization. To this end, the scientific community has deployed more advanced polarization detection experiments, such as ground-based telescope arrays like the Simons Observatory (SO)\citep{Ade2019}, BICEP Array \citep{Hui2018}, CMB-S4 \citep{Abazajian2016}, Ali CMB Polarization Telescope \citep[AliCPT,][]{Li2018}, as well as space-based missions like LiteBIRD \citep{Matsumura2014}, CORE \citep{Delabrouille2018}, and PICO \citep{Hanany2019}. These CMB experiments are designed to achieve multifrequency coverage and high sensitivity. This progress undoubtedly poses greater challenges for data processing, making the exploration of innovative methods to break through traditional data processing techniques essential.

One of the major contaminants in CMB observations is Galactic diffuse foregrounds, which require specialized techniques to remove from the observed data. For CMB polarization, the primary components of these foregrounds are synchrotron and thermal dust emissions, which contribute polarized signals at levels several orders of magnitude higher than the CMB signal across almost all observed frequency bands \citep{BICEP2/KeckCollaboration2015,Krachmalnicoff2016}. Various approaches are employed to analyze multifrequency CMB observations, aiming to separate the CMB signal from foreground contamination. Component separation methods can be broadly categorized into two types: blind methods and non-blind methods. Non-blind methods typically rely on prior knowledge of the emission characteristics of foregrounds. For example, the Commander method \citep{Eriksen2008} utilizes Gibbs sampling to estimate the joint foreground-CMB posterior distribution. In contrast, the ForeGroundBuster method \citep{Stompor2009} employs an analytically derived likelihood function for spectral parameters. However, a key limitation of these methods is their dependence on prior knowledge of foregrounds. When faced with unknown or complex foregrounds, such approaches may introduce biases in the detection of CMB B-modes \citep{ArmitageCaplan2012,Remazeilles2016,Hensley2018}. In contrast, blind source component separation algorithms do not rely on prior knowledge of foregrounds. Instead, they leverage the independent statistical properties of different components to separate signals. Examples of blind methods include spectral matching independent component analysis \citep{Delabrouille2003}, independent component analysis \citep{Baccigalupi2004}, and variants of the internal linear combination (ILC) method \citep[ILC,][]{Eriksen2004,Tegmark2003} \citep[NILC,][]{Delabrouille2009}.

Focusing on foreground contamination removal, deep learning methodologies have been extensively studied in recent years, with Convolutional Neural Networks (CNNs) demonstrating significant potential for reconstructing CMB signals \citep{Wang2022,Casas2022,Yan2023,Casas2023,Yan2024,Sudevan2024,Pal2024,McCarthy2025,Adak2025}. Moreover, these methods have been widely extended to other CMB-related tasks, such as CMB lensing reconstruction \citep{Caldeira2019,Yan2023a,Heinrich2024,Floess2024}, CMB delensing \citep{Yan2023b,Heinrich2024}, and point source processing \citep{Bonavera2021,Casas2022a,Lambaga2025}. However, CNNs are inherently limited to processing flat images within Euclidean space, necessitating the transformation of HEALPix \citep{Gorski2005} spherical sky maps into flat projections. This transformation inevitably introduces artificial boundaries, which can compromise data integrity.
In fact, deep learning methodologies capable of effectively processing HEALPix spherical sky maps remain an active area of research. \cite{Krachmalnicoff2019} extended the one-dimensional CNN approach to HEALPix maps by reorganizing pixels according to HEALPix pixelization properties and performing one-dimensional convolutions. While innovative, this method introduces additional computational complexity. In contrast, DeepSphere \citep{Perraudin2019} proposed a deep learning model using Graph Convolutional Networks (GCNs) tailored for HEALPix sky maps, treating each pixel as a node in a graph structure.  Studies by \cite{Petroff2020} and \cite{Adams2023} indicate that the GCNs method can be effectively applied to the removal of foreground emissions from the CMB temperature data,  potentially addressing the issue of boundary effects.

The Transformer model \citep{Vaswani2017}, as explored by \cite{Liu2021} and \cite{Carlsson2023}, shows promise in processing HEALPix sky maps. However, these studies primarily focus on spherical images encountered in daily life and have not been extensively applied to CMB maps. In this paper, we introduce a novel method based on the Transformer model, referred to as \texttt{TCMB} (Transformer for Cosmic Microwave Background). We utilize this method to remove foreground contamination from CMB polarization data in ground-based observations. Our results demonstrate that \texttt{TCMB} can effectively process HEALPix sky maps and handle regions of arbitrary size, offering a robust and flexible solution for CMB data analysis.

%AliCPT is a ground-based CMB experiment located in the Ali region of Tibet, China. Its primary scientific objective is to detect primordial gravitational waves in the Northern celestial hemisphere. The AliCPT-1 aims to measure CMB polarization signals at two frequencies, 95 GHz and 150 GHz, with observations expected to commence in 2025. Building upon traditional methods, the foreground clean-up strategy for AliCPT-1 has been effectively developed \citep{Ghosh2022,Dou2024,Zhang2024a,Zhang2024}. This work employs the deep learning method \texttt{TCMB} to  remove foreground of CMB polarization for AliCPT-1 experiment.

The paper is organized as follows.  Section \ref{sec:method} provides a comprehensive introduction to the \texttt{TCMB} method.  In Section \ref{sec:mock-data}, we describe the mock data. The results of foreground removal are presented in Section \ref{sec:result}. Finally, we summarize and conclude in Section \ref{sec:conclusions}.

\begin{figure*}
	\centering
	\subfigure[95 GHz]{
		\includegraphics[width=7cm]{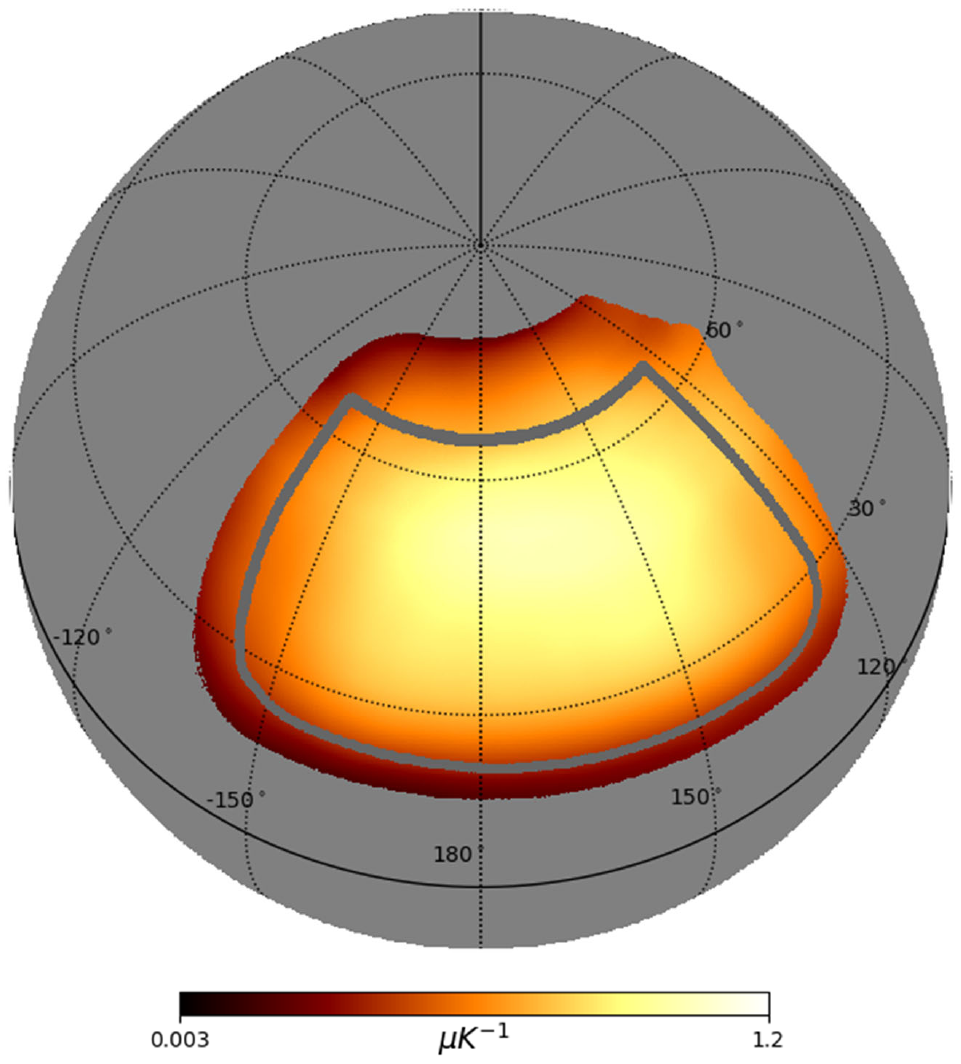}
		%\caption{fig1}
	}
	\vspace{-1mm}
	\hspace{1mm}
	%\quad
	\subfigure[150 GHz]{
		\includegraphics[width=7cm]{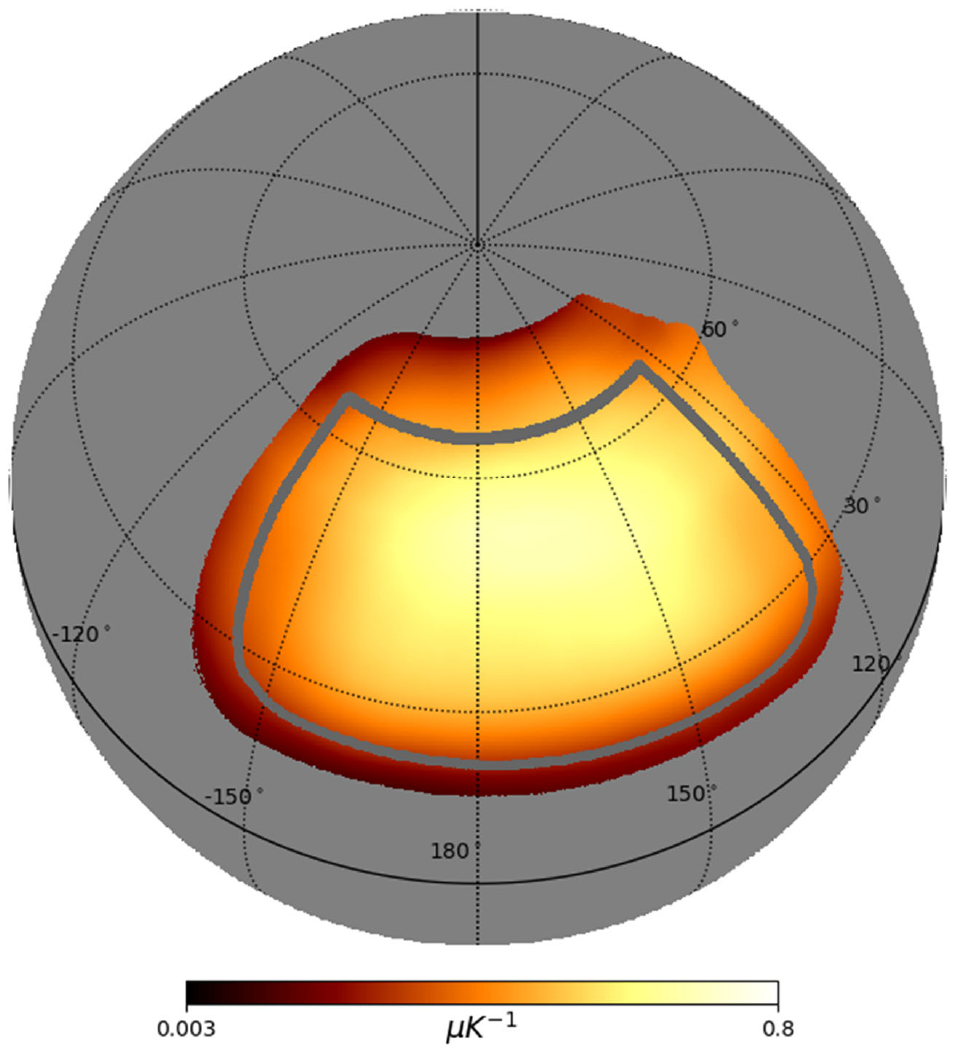}
	}
	\caption{Inverse standard deviation of the polarization noise at frequencies of 95 GHz and 150 GHz, calculated at a resolution of $\rm N_{side}=1024$.The gray line indicates the boundaries of the UNP mask. }
	\label{noise_dist}
\end{figure*}

\section{Method}
\label{sec:method}
Transformer is a relatively novel neural network architecture that primarily employs the self-attention mechanism \citep{Bahdanau2014} to extract intrinsic features \citep{Vaswani2017}, demonstrating significant potential for widespread application in artificial intelligence (AI). Transformer was initially implemented in the domain of natural language processing (NLP), where it demonstrated substantial enhancements in performance \citep{Vaswani2017,Devlin2018}. \cite{Brown2020} pre-trained a large transformer-based model known as GPT-3 (short for Generative Pre-trained Transformer 3), which comprises 175 billion parameters. It demonstrated impressive performance across various natural language processing tasks. Inspired by the significant success of transformer architectures in the field of NLP, researchers have applied this architecture to computer vision (CV) tasks. The ViT model \citep{Dosovitskiy2020} is the first  implementation of the Transformer architecture in the domain of computer vision (CV). In the field of CV, CNNs have long been regarded as fundamental components; however, the transformer architecture is now emerging as a potential alternative to CNNs. To date, the Transformer has demonstrated its efficiency and has shown strong competitive performance compared to CNNs. In this work, we utilize the Transformer to remove the polarized foregrounds of CMB. 

\subsection{Patches}
\label{sec:patch}
The standard Transformer receives as input a 1D sequence of tokens. To handle 2D images, the input image in ViT model is initially segmented, flattened, and subsequently transformed into lower-dimensional linear representations referred to as patch embeddings. Specifically, image $\mathbf{X}\in\mathbb{R}^{H\times W\times C}$ is reshaped into a sequence of flattened 2D patches $\mathbf{X}_p\in\mathbb{R}^{N\times(P^2\cdot C)}$, where ($H$, $W $) is the resolution of the original image, C is the number of channels, ($P$, $P$) is the resolution of each image patch, and $N = HW/P^2$ is the resulting number of patches. These flattened patches are mapped to a $D$ dimensional space through a trainable linear projection layer, and can be represented in matrix form as $f(\mathbf{X}_p)\in\mathbb{R}^{N\times D}$. Thus, we convert image $\mathbf{X}$  into a sequence $f(\mathbf{X}_p)$ that can be used as input for the Transformer, with a sequence length of $N$. Inspired by the ViT model, we transform the HEALPix map into sequence data suitable for processing by Transformer, based on the pixelization characteristics of the HEALPix representation.

HEALPix \citep{Gorski1999,Gorski1999a,Gorski2005} is a general structure for the pixelization of data on the sphere.  It offers an authentic curvilinear partitioning of the sphere into quadrilaterals of exactly equal area, distinguished by their varied geometrical shapes. Specifically, HEALPix partitions the sphere into $12\times {\rm \texttt{NSIDE}}^2$ equal-area grids, where \texttt{NSIDE} represents the resolution of the grid. The boundaries for \texttt{NSIDE}=1 define the $12$ base-resolution pixels and higher-order pixelisations are defined by their regular subdivision. \texttt{NSIDE} describes the number of subdivisions along the edges of a base-resolution pixel required to achieve the desired high-resolution partition. HEALPix support two different numbering schemes for the pixels, \texttt{RING} scheme and \texttt{NESTED} scheme. The former is suited for spherical harmonic analysis. The latter refers to a pixel nesting arrangement that conveniently elucidates the structural configuration of the pixels and the relationships among adjacent pixels. In the sky map of the \texttt{NESTED} scheme, the inherent nesting pattern of pixels facilitates a natural division of the sky map into multiple equal-area patches. Inspired by the ViT model, we aggregate these patches into a sequence data suitable for processing by Transformer architecture.

\begin{figure}
	\centering
	\includegraphics[width=0.7\hsize]{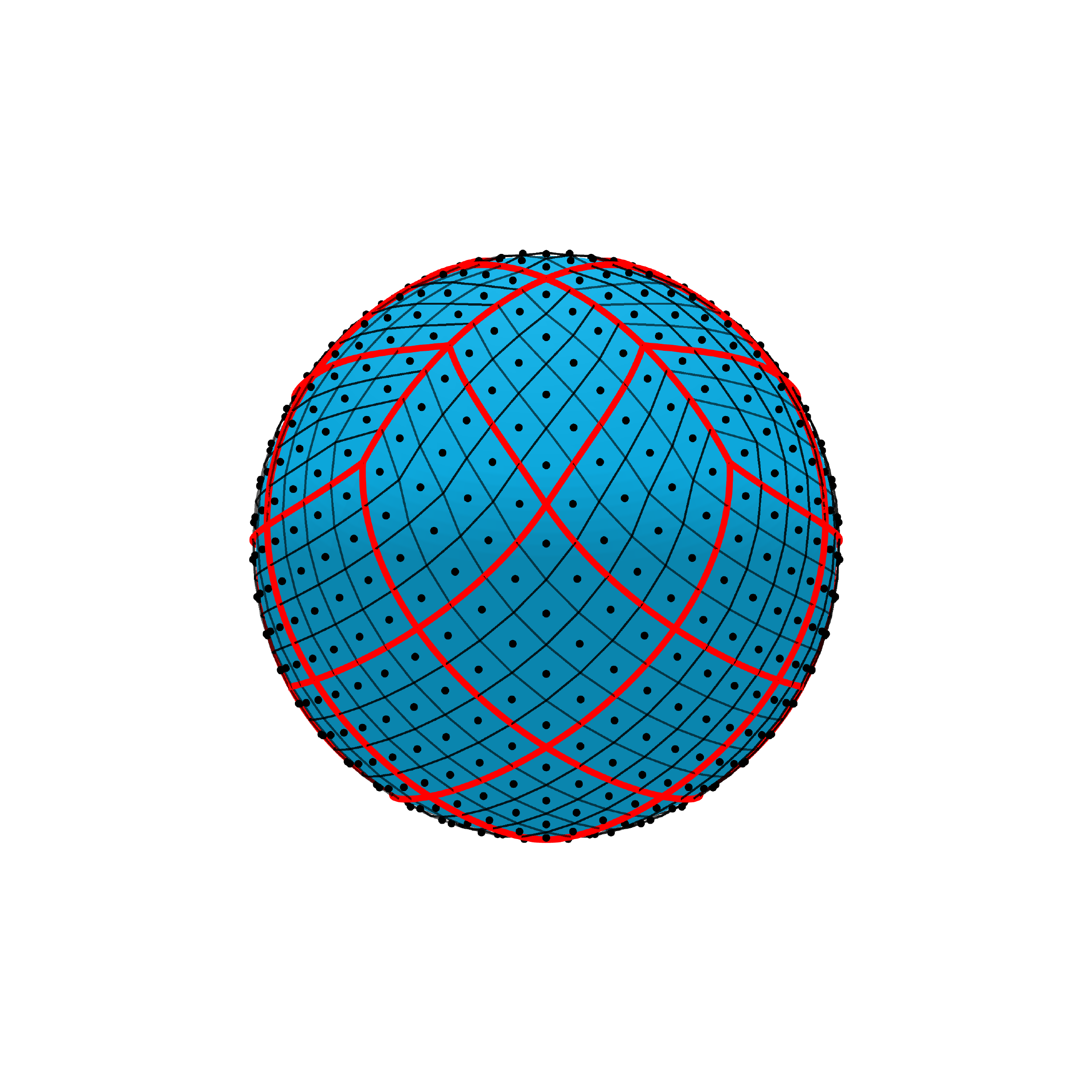}
	\caption{Orthographic view of the HEALPix partition of the sphere. The resolution parameter of the map is set to \texttt{NSIDE}=8. The black points and lines represent the center positions and boundaries of the map pixels, respectively. The red lines denote the boundaries of the defined patch, which has its size parameter set to ${\rm \texttt{NSIDE}_P}=2$.}  
	\label{healpix_sphere}
\end{figure}
\begin{figure*}
	\centering
	\includegraphics[width=1\hsize]{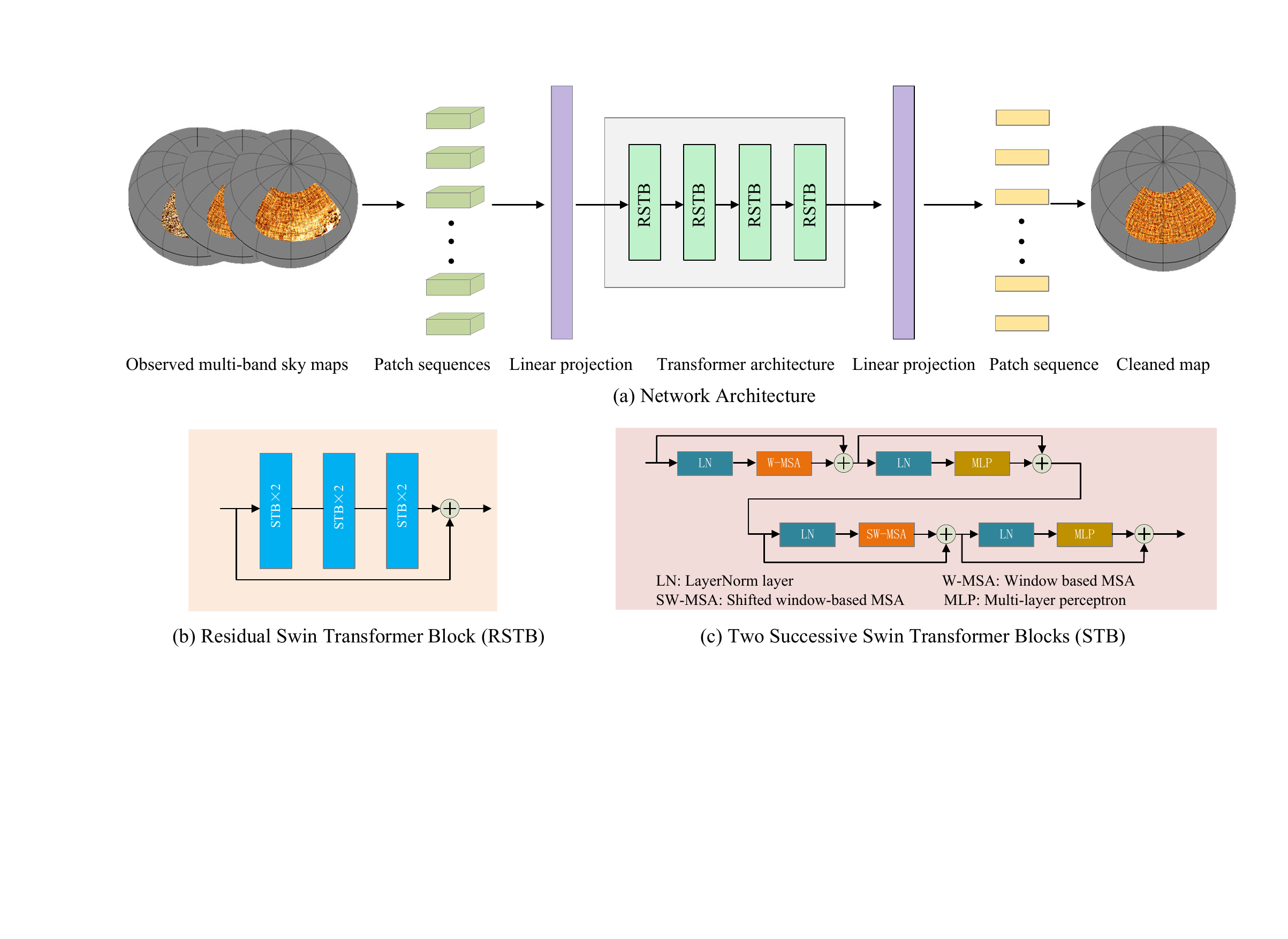}
	\caption{The network architecture for \texttt{TCMB}.}  
	\label{fig:network}
\end{figure*}
We consider a full-sky map with  resolution parameter \texttt{NSIDE}. First, we define the parameter ${\rm \texttt{NSIDE}_P}$ to determine the size of the patch, where ${\rm \texttt{NSIDE}_P}$ needs to be a power of two and ${\rm \texttt{NSIDE}_P} \leq $ \texttt{NSIDE}.
Subsequently, the entire full-sky map in \texttt{NESTED} scheme can be naturally segmented into multiple patches, with the total number of patches being ${\rm 12\times \texttt{NSIDE}_P^2}$ and the number of pixels in each patch being ${\rm (\texttt{NSIDE}/\texttt{NSIDE}_P)^2}$. Figure \ref{healpix_sphere} illustrates an example in which the full-sky map with \texttt{NSIDE}=8 is partitioned into multiple patches, with a parameter of ${\rm \texttt{NSIDE}_P}=2$. In the context of analyzing a partial sky, one can employ a mask to identify and retain patches that correspond to the observational coverage, while excluding those that are outside this coverage. This methodology facilitates a significant reduction in the number of patches and, as a result, optimizes the use of computational resources. Finally, we arrange the patches according to their latitude, thereby obtaining sequence data suitable for processing by the Transformer.

Given a full-sky map $\mathbf{Y}\in\mathbb{R}^{{\rm \texttt{Npix}}}$, it can be partitioned into multiple patch sequence data $\mathbf{Y}_p\in\mathbb{R}^{N\times {\rm \texttt{Npix}_P}}$ based on its nested characteristics. Here, $N$, \texttt{Npix}, and $\texttt{Npix}_P$ represent the number of patches, and pixels in the full-sky map and pixels in a patch, respectively. If we consider the multi-band sky maps, denoted as $\mathbf{Y}\in\mathbb{R}^{F\times {\rm \texttt{Npix}}}$, then the corresponding patch sequence can be represented as $\mathbf{Y}_p\in\mathbb{R}^{N\times F \times {\rm \texttt{Npix}_P}}$, where $F$ represent the number of frequency bands. Subsequently, we apply a linear projection layer (or one-dimensional convolution layer)  $f(\cdot)$ to generate the input data $f(\mathbf{Y}_p)\in\mathbb{R}^{N\times ({\rm \texttt{Npix}_P} * F*c)}$ suitable for processing by the Transformer, where $c$ is a constant. In this work, we set the hyperparameter $c=128$. The parameters involved in this  operation ($f(\cdot)$) are acquired through learning. The aim of this operation is to enable the transformation of map patches into vector representations, referred to as patch embeddings (tokens), thereby enabling the Transformer model to process map as sequences of tokens using a transformer-based methodology.

\subsection{Self-attention mechanism}
The core component of the Transformer architecture is the self-attention (SA) mechanism, which can determine the importance of a single token (patch embedding) relative to other tokens in the sequence by computing the relevance between that token and the other tokens \citep{Vaswani2017}.  This mechanism effectively models long-range dependencies, which has resulted in the widespread recognition of the SA mechanism as a key contributor to the success of Transformer models.

Suppose the input data is a set of tokens $\mathbf{I} = (\mathbf{i_1},\mathbf{i_2},...,\mathbf{i_N}) \in\mathbb{R}^{N \times D}$, where $N$ and $D$ is the number of tokens and token dimension. The objective of SA is to capture the interactions among all $N$ tokens by encoding each token in relation to the global contextual information. That is done by defining three learnable weight matrices ($\mathbf{W}^Q  \in\mathbb{R}^{D \times D_q},\mathbf{W}^K \in\mathbb{R}^{D \times D_k},\mathbf{W}^V \in\mathbb{R}^{D \times D_v}$) to transform input $\mathbf{I}$ into Queries ($\mathbf{Q}$), Keys ($\mathbf{K}$), and Values ($\mathbf{V}$):
\begin{equation}
	\label{eq:SA}
	\mathbf{Q} = \mathbf{I} \mathbf{W}^Q, \mathbf{K} = \mathbf{I} \mathbf{W}^K, \mathbf{V} = \mathbf{I} \mathbf{W}^V,
\end{equation}
where $D_q=D_k=D_v$ and are the hidden dimension.
The corresponding attention matrix $\mathbf{A} \in\mathbb{R}^{N \times N}$ can be expressed as follows:
\begin{equation}
	\label{eq:AM}
	\mathbf{A} = {\rm softmax}\left(\frac{\mathbf{QK}^T}{\sqrt{D_q}} \right),
\end{equation}
where softmax is normalized exponential function. Then the SA is performed as follows
\begin{equation}
	\label{eq:SA_o}
	\mathbf{O} = {\rm SA}\left(\mathbf{I} \right) = \mathbf{AV},
\end{equation}
where $\mathbf{O} \in\mathbb{R}^{N \times D_v}$ is the output of SA.

The limited capacity of a single SA module often leads to its focus on only a few positions, potentially overlooking other important positions. In practice, the implementation of multi-head SA (MSA) is predominant. MSA employs the parallel stacking of SA blocks (heads), thereby enhancing the efficacy of the self-attention layer. The idea of MSA is to stack multiple SA blocks in parallel, and the concatenated outputs are fused by a projection matrix $W^O$. In this task, the number of heads is set to 6. Each head has its own learnable weight matrices denoted by $\{\mathbf{W}^Q_j, \mathbf{W}^K_j, \mathbf{W}^V_j\}$, where $j = 0,1,...,h-1$ and $h$ denotes total number of heads in MSA block. The output $\mathbf{O}_j$ of  $j$-th head SA is given by formula \ref{eq:SA_o}. The output of MSA can write,
\begin{equation}
	\label{eq:MSA}
	\mathbf{O}_{\rm MSA} = [\mathbf{O}_0, \mathbf{O}_1, ..., \mathbf{O}_{h-1}]W^O.
\end{equation}

\subsection{Network architecture}

The network architecture employed in this work is illustrated in Figure \ref{fig:network}. As outlined in Section \ref{sec:patch}, the masked multiband observation maps are initially partitioned into multiple patches. Subsequently, these patches are mapped to patch embedding sequence through a linear layer, which are then ultimately input into the Transformer module.

We use Swin Transformer block (STB) \citep{Liu2021a} as the fundamental unit of Transformer module.  Two successive STBs are presented in the panel (c) of Figure \ref{fig:network}. Each STB is comprised of a LayerNorm (LN) layer, a MSA module, a residual connection, and a  multi-layer perceptron (MLP) employing GELU non-linearity. This multi-layer NLP block are composed of an input layer, a hidden layer, and an output layer. Both the input layer and the output layer are configured with ${{\rm \texttt{Npix}_P * F*c}}$ neurons, while the number of neurons in the hidden layer is set to ${4{\rm \texttt{Npix}_P} * F*c}$.  Here, ${\rm \texttt{Npix}_P}$, $F$, and $c$ are consistent with those defined in Section \ref{sec:patch}.  Different from the conventional MSA module, the first block uses window-based MSA (W-MSA), and the second block shifts those windows and applies window-based MSA again (SW-MSA). W-MSA initially partitions tokens into multiple windows, with each window containing several non-overlapping tokens and performing SA calculations independently. Since the number of windows is significantly smaller than the total number of tokens, W-MSA effectively reduces computational complexity compared to MSA. However, there is no interaction between different windows. To facilitate information exchange between distinct windows, the second block employs the SW-MSA module. SW-MSA adjusts the window allocation through a sliding mechanism, ensuring that the newly defined windows contain tokens that differ from those in the first block. After the reallocation of windows, self-attention is performed independently on each window once again. Consequently, the two STBs not only significantly alleviate computational burdens but also enable information exchange among all tokens, thereby enhancing computational efficiency. Further detailed information can be found in reference \citep{Liu2021a}.

The design of the Transformer architecture is based on reference \citep{Liang2021}, consisting of four layers of residual swin transformer block (RSTB), with each RSTB layer composed of  three sets of two successive STBs. The output of the Transformer is subsequently transformed through a linear mapping to obtain the target sky map. The network model constructed in this work involves a total of 173 million parameters during the training process.  This scale reflects the high complexity of the model and indicates that substantial computational resources are required for training, which imposes corresponding demands on optimization efficiency and hardware configuration.

\section{Mock Data Sets}

\subsection{Simulation}
\label{sec:mock-data}
Deep learning methods rely heavily on large datasets to optimize models effectively. In this section, we describe the process for generating the dataset utilized in this study.

%AliCPT is a ground-based CMB experiment located in the Ali region of Tibet, China. Its primary scientific objective is to detect primordial gravitational waves in the Northern celestial hemisphere. The AliCPT-1 aims to measure CMB polarization signals at two frequencies, 95 GHz and 150 GHz, with observations expected to commence in 2025. The observational sky patch of the AliCPT-1 instrument is located in the Northern Hemisphere, with central coordinates at R.A.=$180^{\circ}$ and decl.=$30^{\circ}$, approximating an area that covers 17\% of the sky. Figure \ref{noise_dist} show observational sky patch of the AliCPT-1 at ``20 $\mu$K" mask. ``20 $\mu$K" mask is produced by removing any pixel with the noise standard deviation above 20 $\mu$K at $\rm N_{side}$=1024 for the 150GHz channel and the accepted sky fraction is about 10\%. ``UNP" mask further remove the sky above a decl. of 65° to mitigate foreground contamination, the accepted sky fraction is about 6.7\%. In this work, we adopt``UNP" mask. Besides the data in the 95 and 150 GHz bands of AliCPT-1, we will integrate data from the Planck experiment to construct multi-frequency observational datasets and simulate data in the Planck High Frequency Instrument (HFI)’s four bands at 100, 143, 217, and 353 GHz.

We adopt a ground-based CMB polarization experiment as the baseline for generating our dataset. Specifically, we use the AliCPT-like experiment \citep{Li2018,Ghosh2022,Dou2024,Zhang2024a,Zhang2024} as our instrument configuration baseline, as it is one of the telescopes currently in operation and is capable of hosting large detector arrays to provide highly precise measurements of CMB polarization in the short near future. The AliCPT-1 experiment aims to measure CMB polarization signals at two frequencies, 95 GHz and 150 GHz. The observational sky patch for AliCPT-1 is located in the Northern Hemisphere, centered at R.A. = 180° and decl. = 30°, covering approximately 17\% of the sky.

Figure \ref{noise_dist} shows the observational sky patch of AliCPT-1 using the ``20 $\mu$K" mask. This mask is generated by excluding pixels with a noise standard deviation above ``20 $\mu$K" at $\rm N_{side}$=1024 for the 150 GHz channel, resulting in an accepted sky fraction of about 10\%. The ``UNP" mask further removes the sky above a declination of 65° to mitigate foreground contamination, reducing the accepted sky fraction to approximately 6.7\%. In this work, we adopt the ``UNP" mask.
In addition to the data from the 95 GHz and 150 GHz bands of AliCPT-1, we combine data from the Planck experiment to construct multi-frequency datasets, by simulating data in the Planck High Frequency Instrument (HFI)’s four bands: 100 GHz, 143 GHz, 217 GHz, and 353 GHz.

The CMB simulation are Gaussian realizations generated from the power spectra. Specifically, we initially employ the \texttt{CAMB}\footnote{https://github.com/cmbant/CAMB} \citep{Lewis2000} software package to compute the CMB power spectra and lensing power spectrum within the framework of $\Lambda$ cold dark matter cosmology ($\Lambda$CDM). The tensor-to-scalar ratio $r$ is fixed to $0.023$. Other cosmological parameters relevant to the $\Lambda$CDM framework include \(H_0\), \(\Omega_{b}h^2\), \(\Omega_{c}h^2\), \(\tau\), \(A_s\), and \(n_s\). The best-fit value and corresponding standard deviations for these parameters are derived from the Planck 2018 parameters \citep{PlanckCollaboration2020a}. The lensing effects are added by the \texttt{LensPyx} \citep{Reinecke2023} package. 

In the foreground simulation, we take into account various components of the Galactic diffuse foreground polarization, including thermal dust, synchrotron radiation, and anomalous microwave emission (AME). The foreground maps are generated from the \texttt{PySM3} \footnote{https://github.com/galsci/pysm/tree/main} package \citep{Thorne2017,Zonca2021}. The simulation of  synchrotron radiation is performed by extrapolating template maps based on a parametric model derived from the \texttt{PySM} s1 model. The thermal dust and AME derived from the \texttt{PySM} d1 model and a2 model. 

\begin{table}
	\begin{center}
		\centering
		\caption{Frequencies and instrumental specifications.}\label{table_1}
		\begin{tabular}{ c|c|c}
			\hline\hline
			
			Experiment & AliCPT & Planck HFI  \\
			\hline
			Frequency (GHz) & 95\hspace{0.3em} 150 &100\hspace{0.3em} 143\hspace{0.3em} 217\hspace{0.3em} 353 \\
			\hline
			Beam size (arcmin) & 19\hspace{0.3em} 150 & 9.7\hspace{0.3em} 7.3\hspace{0.3em} 5.0\hspace{0.3em} 4.9\\
			
			\hline\hline
		\end{tabular}
	\end{center}
\end{table}
The instrumental noises for AliCPT-1 are Gaussian realizations generated from the noise standard deviation maps of AliCPT-1 bands. The noise of AliCPT-1 exhibits anisotropy, with the distribution of noise levels presented in Figure \ref{noise_dist}. This noise level is derived from four years of observational data and the the total integrated observation time is projected to be ``48 module*years". The harmonic mean of the noise level are 3.2 $\mu$K-arcmin for 95 GHz and 4.9 $\mu$K-arcmin for 150 GHz. Here, we utilize the half-mission noise level, which is enhanced by a factor of $\sqrt{2}$ in comparison to the full-mission noise level. Planck half-mission noise maps are based on the FFP10 noise simulations and downloaded from the Planck Legacy Archive (PLA) \footnote{https://www.cosmos.esa.int/web/planck/pla}. For the half-mission noise simulation, PLA provides only 300 noise realizations, resulting in a total of 600 noise maps when considering two half-mission. To augment the number of noise maps, we employ a linear combination method. The specific steps are as follows: (1) A set of 10 random numbers ($n_i$, $i=0,1,...,9$) is generated in accordance with a uniform distribution, to compute the associated weights. The weight ($w_i$) is determined through the normalization of the respective random number by the sum of all generated values, as expressed by the equation: $w_i=n_i/(\sum_{i=0}^{9}n_i)$; (2) Randomly select 10 noise maps from the total of 600 noise maps; (3) Multiply these 10 maps by their respective weights and sum them to obtain a new noise map.

In the process of constructing a deep learning training dataset, effective expansion of the data sample space is required to enhance the statistical diversity and coverage of the data, thereby improving the generalization capability of the network model. Specifically, in each realization, cosmological parameters are randomly sampled from a uniform distribution within a range of five standard deviations, and the corresponding CMB sky map for each sample is treated as an independent Gaussian random realization.

For foreground simulation, to increase the diversity of foreground samples, we employ a parameter perturbation method to model the foreground templates. This includes introducing a 10\% Gaussian random error to the amplitude parameters and a 5\% Gaussian random error to the spectral indices. This perturbation strategy effectively broadens the coverage of the sample space while preserving the fundamental physical characteristics of foreground radiation. In each realization, we first apply independent random perturbations to the amplitude and spectral index parameters of the foreground template. Specifically, a Gaussian random variable with zero mean and a standard deviation of 0.1 is generated and multiplied by the amplitude template map to produce a perturbation component. This component is then added to the original amplitude template to generate the perturbed amplitude map. Similarly, for the spectral index parameter, a Gaussian random variable with zero mean and a standard deviation of 0.05 is generated and multiplied by the spectral index template sky map to obtain a perturbation term. This term is then superimposed onto the original spectral index map to form the perturbed spectral index map. Finally, the foreground emission is simulated based on the perturbed amplitude and spectral index maps.

We simulate a total of 50,000 sky maps across six frequency bands to construct the training dataset, with all sky maps uniformly set to a spatial resolution of \texttt{NSIDE}=512. The test set comprises independently generated 2,000 sky maps, whose sampling methodology for cosmological parameters and foreground model parameters remains consistent with the training set, both employing independent sampling. The noise data used in the test set were directly sourced from the Planck Legacy Archive (PLA) without any weighting processing applied, thereby preserving the authenticity and originality of the noise.

\subsection{Dataset and Training}
The inputs to the network model consist of beam-convolved observational Q or U maps derived from all six frequency bands. The desired output of the network is a beam-convolved CMB Q or U map with beam size of  95 GHz, plus noise at 95 GHz. In reference \citep{Yan2024}, we highlighted the challenges faced by deep learning methods in effectively separating the CMB polarization from noise on the map level. Therefore, in this study, our training objective continues to take into account data that includes both the CMB and noise, while concentrating the network model's efforts on the removal of the foreground. We select the noise in the 95 GHz frequency band as the desired output noise of the network, primarily due to the lower noise level associated with this band compared to others. A detailed discussion on the selection of frequency bands can be found in reference \citep{Yan2024}.

\begin{figure*}
	\centering
	\subfigure{
		\includegraphics[width=16cm]{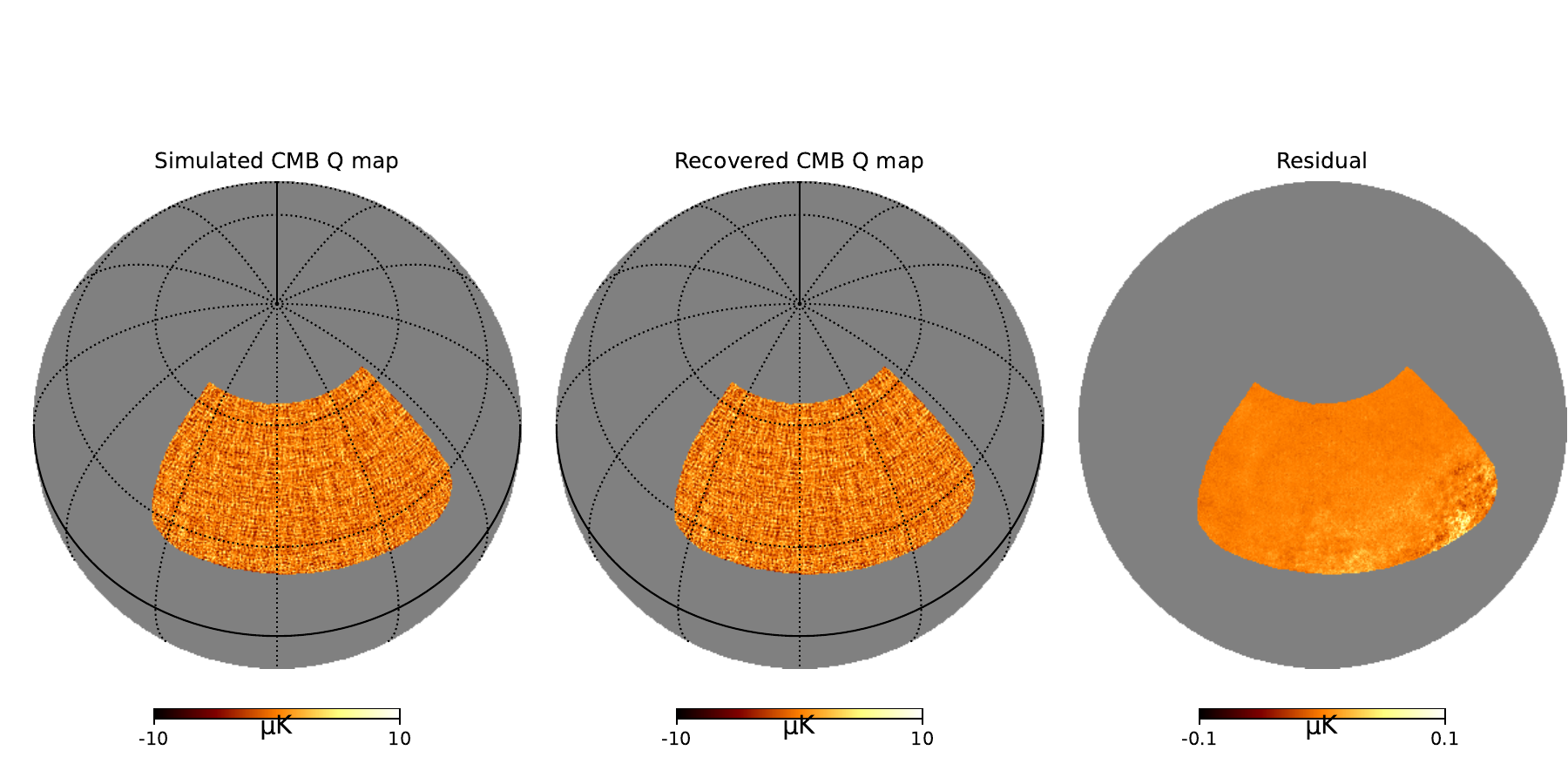}
		%\caption{fig1}
	}
	%\vspace{-1mm}
	%\hspace{1mm}
	%\quad
	\subfigure{
		\includegraphics[width=16cm]{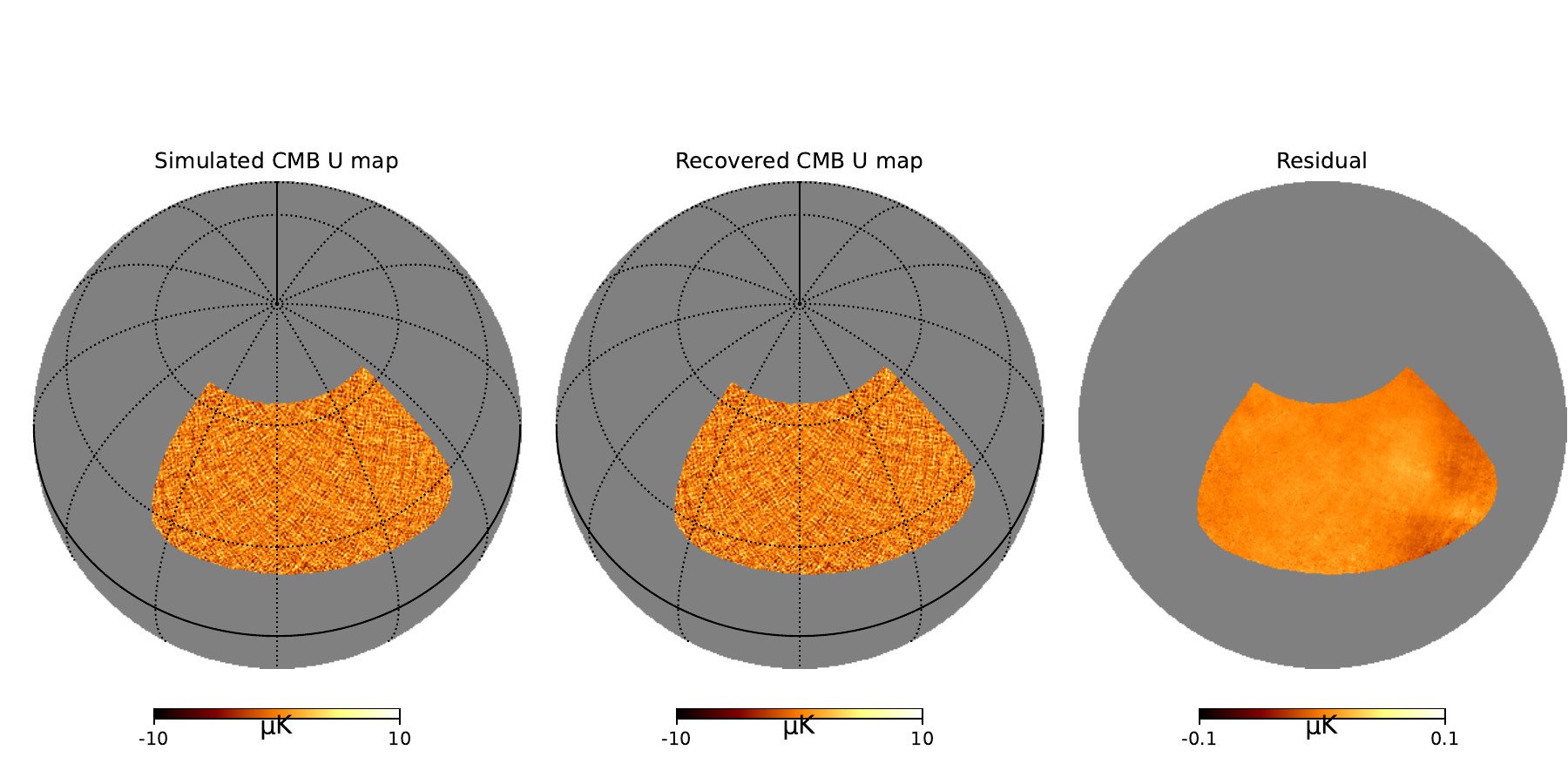}
	}
	\caption{The results of foreground removal using \texttt{TCMB} method at map level. The upper panels illustrate the recovery of the Q map, while the lower panels depict the recovery of the U map. The simulated maps are a beam-convolved CMB map plus a noise map at a frequency of 95 GHz. The recovered maps correspond to the noisy CMB maps that were recovered using the \texttt{TCMB} method from multi-band observational data. The residual maps indicate the differences between the recovered map and the target map.}
	\label{fig:recovered-maps}
\end{figure*}
\begin{figure*}
	\centering
	\includegraphics[width=1\hsize]{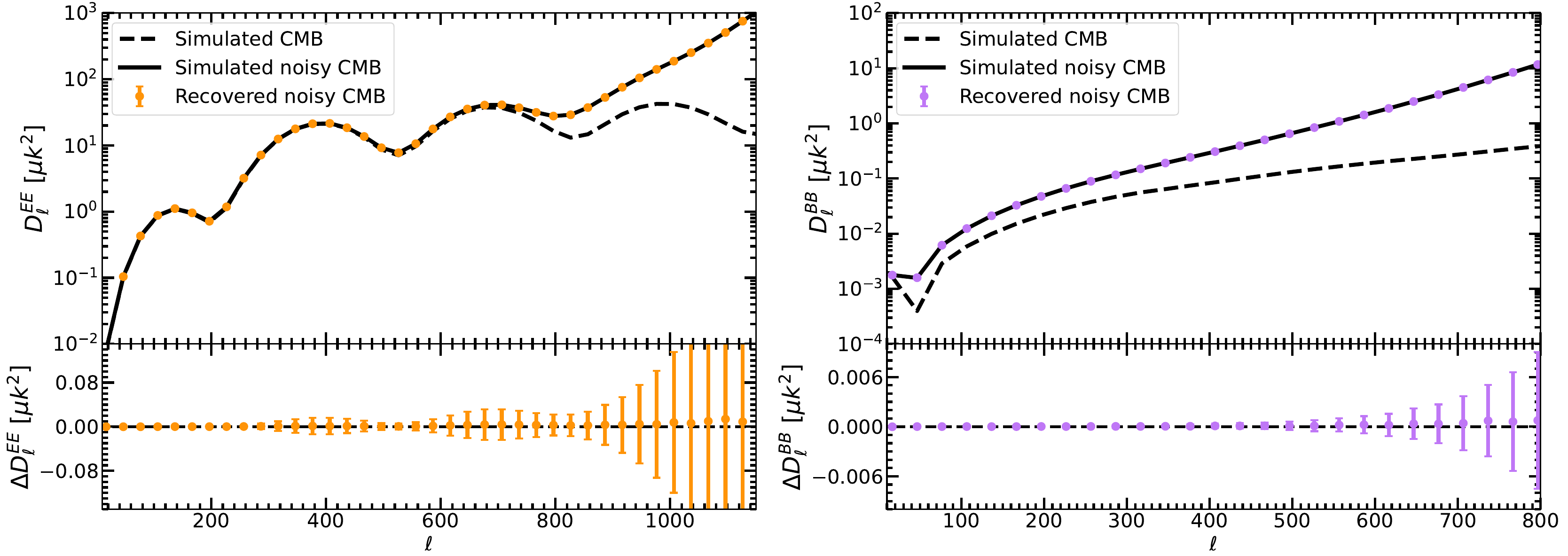}
	\caption{Recovered power spectra of the E-mode (EE) and B-mode (EE) of the CMB, incorporating contributions from noise. The simulated power spectra were calculated from the beam-convolved CMB map plus a noise map at a frequency of 95 GHz. The recovered power spectra were derived from the noisy CMB maps, which were obtained through the application of the \texttt{TCMB} method on multi-band observational data. $\Delta D_{\ell}$ represents the difference between the recovered power spectrum and the target power spectrum, $\Delta D_{\ell}= D_{\ell ,\rm recovered} - D_{\ell, \rm true}$. The black dashed lines represent the power spectra calculated using the pure CMB map. The length of each $\ell$ bin is set to be 30. }  
	\label{fig:noisy_ps}
\end{figure*}
First, we preprocess the observational sky maps, utilizing a method derived from the ILC. The ILC method requires the observed data to be deconvolved to a common beam and resolution \citep{Delabrouille2009,Remazeilles2011}. Here, we employ this method for preprocessing the observed data \citep{Adak2025}. We consider multi-frequency observational sky maps $M^{\rm obs,\nu}(p)$, wherein each frequency band is characterized by distinct instrumental beam profiles. The index $\nu$ and $p$ denote frequency band and pixel, respectively. The observed maps convolves/deconvolutes to the same resolution in harmonic space:
\begin{equation}
	\label{eq:MS}
	M_{\ell m}^\nu=\frac{b_\ell^c}{b_\ell^\nu}M_{\ell m}^{\mathrm{obs},\nu},
\end{equation}
where, $M_{\ell m}^{\mathrm{obs},\nu}$ is harmonic coefficients of $M^{\rm obs,\nu}(p)$. $b_\ell^\nu$ and ${b_\ell^c}$ denote the beam window function associated with each frequency band and the common beam window function, respectively. In order to maintain a correspondence with the desired output of the network, the observational maps are also reconvolved with a common beam of the 95 GHz. However, it is important to note that in our tests, we found that this data preprocessing method did not have a significant impact on the results compared to the case without preprocessing. If there are substantial errors in the instrumental beam, this preprocessing may be deemed unnecessary.

After the pre-processing of the multi-band observed sky maps, it is necessary to partition the maps into multiple patches. The sky map has \texttt{NSIDE}=512, and we employ a parameter of ${\rm \texttt{NSIDE}_P}=64$ for the partitioning. Consequently, the full-sky map can be divided into $12 \times 64^2 = 49152$ patches, with each patch containing 64 pixels. In our tests, slight variations in ${\rm \texttt{NSIDE}_P}$, such as increasing it to ${\rm \texttt{NSIDE}_P}=256$ or decreasing it to ${\rm \texttt{NSIDE}_P}=32$, have a negligible impact on the results. We utilize a mask to retain the patches covered by observations while excluding those that are not observed. The effective number of patches is 3494. To facilitate the computations in the Transformer model, we employ zero padding to expand the number of patches to 3600. This procedure is solely intended to facilitate computation, and the padded patches will not be considered in subsequent data analyses.

During the training process of the neural network model, the optimization of model parameters is achieved by minimizing the loss function. The loss function serves to quantify the discrepancy between the network's predictions and the target. In this study, we have chosen to employ the mean absolute deviation (L1) loss function as our optimization criterion. To optimize network model's performance, we employ the Adam optimizer, as proposed by \cite{Kingma2014}, initializing the learning rate at $3\times 10^{-4}$ and gradually decreasing it over the course of the iterations until it reaches a final value of $1\times 10^{-7}$. The training process consists of approximately 200 epochs, with a batch size of 24, and is conducted on eight NVIDIA RTX 4090 GPUs. On average, the training process for a single network model takes approximately 70 hours to complete.

\section{Results}
\label{sec:result}
%In this section, we will utilize the proposed method, \texttt{TCMB}, to perform foreground subtraction on the CMB polarization data, and we will present the results of the foreground subtraction.
In this section, we employ the proposed method, \texttt{TCMB}, to conduct foreground removal on CMB polarization data and present the outcomes of this analysis.

\begin{figure*}
	\centering
	\includegraphics[width=1\hsize]{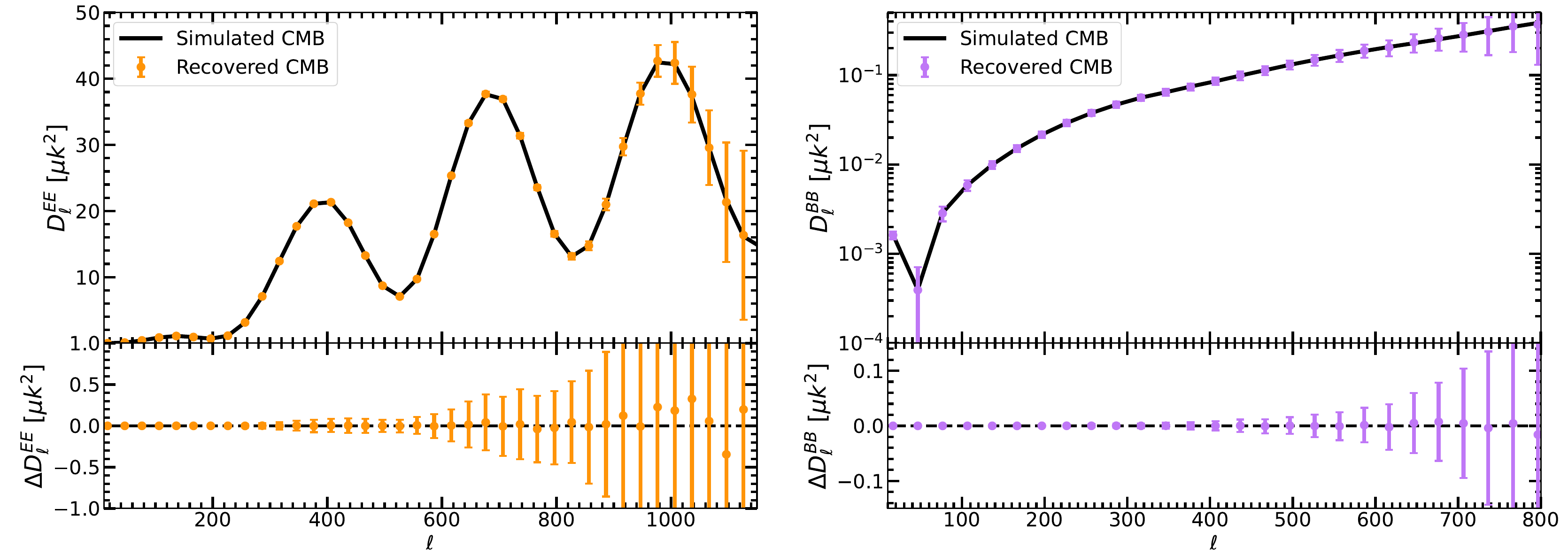}
	\caption{CMB EE and BB power spectra estimated from \texttt{TCMB} foreground cleaning pipelines. The black line represents the power spectra calculated from the pure CMB map with UNP mask.}  
	\label{fig:recov_ps}
\end{figure*}
\subsection{Recovered CMB map}
\label{sec:recoveredCMB}
We first present the results of foreground removal at the map level. As illustrated in Figure \ref{fig:recovered-maps}, the recovered CMB polarization map, including noise components, shows a high degree of consistency with the true map. The information remaining in the residual map is minimal, indicating that the diffuse foreground has been effectively and cleanly removed. To quantitate the effectiveness of foreground removal, we employed the calculation of the mean absolute deviation (MAD):
\begin{equation}
	\label{eq:MSA_recov}
	\sigma_{\rm MAD}=\frac{1}{N}\sum_{i}^{N}|X_i-Y_i|,
\end{equation}
where $N$ is the number of pixels, $X$ and $Y$ represent the predicted and real sky maps. The learning objective of the network model is to minimize the MAD value as much as possible, aiming for it to approach zero. In the test set, the values of the MAD are $\sigma_{\rm MAD}^Q = 0.0081 \pm 0.0036$ $\mu$K for recovery of CMB Q map and $\sigma_{\rm MAD}^U = 0.0073 \pm 0.0039$ $\mu$K for recovery of CMB U map. These values are approximately 1/200 of the CMB polarization signal, indicating that the residual foreground is quite limited. This further validates the high efficiency of our method in effectively removing foreground contamination.

\subsection{Recovered CMB power spectra}
After obtaining the foreground-removed CMB map, we utilized the \texttt{NaMaster}  package \citep{Alonso2019}  to compute the power spectrum. The $C2$ function, as implemented in \texttt{NaMaster}, was utilized to apodize the UNP mask prior to the computation of the power spectra, employing apodization scales of $4^{\circ}$. Figure \ref{fig:noisy_ps} illustrates the power spectrum of the CMB map reconstructed by the network model, which includes instrumental noise. The error bars are derived from statistical results obtained from the test set and represent the standard deviation between the power spectrum calculated from the target CMB map and that computed from the CMB map reconstructed by the network model.

For the E-mode power spectrum, we can see that the CMB EE spectrum reconstructed by the network model is in high agreement with the target power spectrum. However, due to the presence of instrumental noise in the recovered CMB map, the power spectrum exhibits a significant deviation from the true CMB EE power spectrum at smaller scales. The BB power spectrum is predominantly dominated by instrumental noise. We can see that the BB power spectrum reconstructed by the network model is also in strong agreement with the target power spectrum. These results indicate that the diffuse foreground can be effectively removed, thereby preserving the CMB and its associated noise information in full.

To obtain a cleaner CMB power spectra, it is necessary to apply debiasing techniques of noise to the recovered CMB maps by network. Here, we employ two half-mission maps for cross-correlation analysis to facilitate the debiasing of noise.  We can segment the observed time-ordered data (TOD) from the entire survey mission into two halves, corresponding to two half-mission datasets. These two data subsets contain the same CMB signal and foregrounds, but their noise is uncorrelated. By calculating the cross-correlation power spectrum between the two half-mission maps, we can effectively reduce the noise present in the power spectrum. We utilize the network model to perform individual foreground removal on each half-mission map, resulting in two foreground-cleaned CMB sky maps. Subsequently, we calculate their cross-correlation power spectrum.

The power spectra after noise debiasing is presented in Figure \ref{fig:recov_ps}. The error bars represent the $1\sigma$ statistical uncertainty derived from the test set, with this uncertainty primarily originating from instrument noise. Notably, at smaller scales, the uncertainty introduced by instrument noise becomes significantly pronounced. We can see that the reconstructed CMB EE and BB power spectra are consistent with the true values, indicating that \texttt{TCMB} foreground cleaning pipelines is highly effective.

\subsection{Comparison with CNN method}
\label{sec:CNN}
%References \citep{Yan2023,Yan2024} proposed a method for CMB foreground subtraction based on convolutional neural networks (CNN), known as \texttt{CMBFSCNN}. Here, we present the results using the \texttt{CMBFSCNN} method and provide a comparison between the \texttt{TCMB} method and the \texttt{CMBFSCNN} method. 

References \citep{Yan2023,Yan2024} have demonstrated a method for CMB foreground removal based on convolutional neural networks (CNNs), known as \texttt{CMBFSCNN}. Here, we present a comparison between the results of the \texttt{CMBFSCNN} method and those of the \texttt{TCMB} method. 

To ensure rigorous fairness and comparability in our experimental evaluation, we conduct a performance comparison between Transformer-based and CNN-based methodologies using an identical test set configuration. It is crucial to highlight that while the the evaluation framework remained consistent, the two architectures exhibit fundamental differences in training data requirements and model complexity  due to their intrinsic structural characteristics - specifically, the global self-attention mechanism inherent in Transformers versus the localized receptive fields characteristic of CNNs. Our comprehensive hyperparameter optimization process, incorporating grid search and cross-validation techniques, determin that the \texttt{CMBFSCNN} architecture achieves optimal performance with merely 1,000 training samples, whereas the \texttt{TCMB} model requires an extensive 50,000-sample training set. Furthermore, the two models exhibit distinct parameter scales. The \texttt{CMBFSCNN} network contains 6.31 million trainable parameters, representing only 3.7\% (approximately 1/27th) of the \texttt{TCMB} network's parameter count. This parameter efficiency grants the \texttt{CMBFSCNN} architecture notable advantages in terms of model simplicity and resource efficiency. Most significantly, our experiments revealed striking differences in computational demands. The \texttt{TCMB} implementation, utilizing Transformer architecture, exhibited substantially greater resource requirements during training. Under our experimental configuration (8×NVIDIA RTX 4090 GPUs), a single \texttt{TCMB} training cycle required approximately 70 hours to complete. In stark contrast, the \texttt{CMBFSCNN} model demonstrated remarkable computational efficiency, achieving full training convergence in just 12 hours while utilizing only 2 GPUs. These empirical results unequivocally demonstrate the \texttt{CMBFSCNN} architecture's superior computational efficiency throughout the training process.

\begin{figure}
	\centering
	\includegraphics[width=1\hsize]{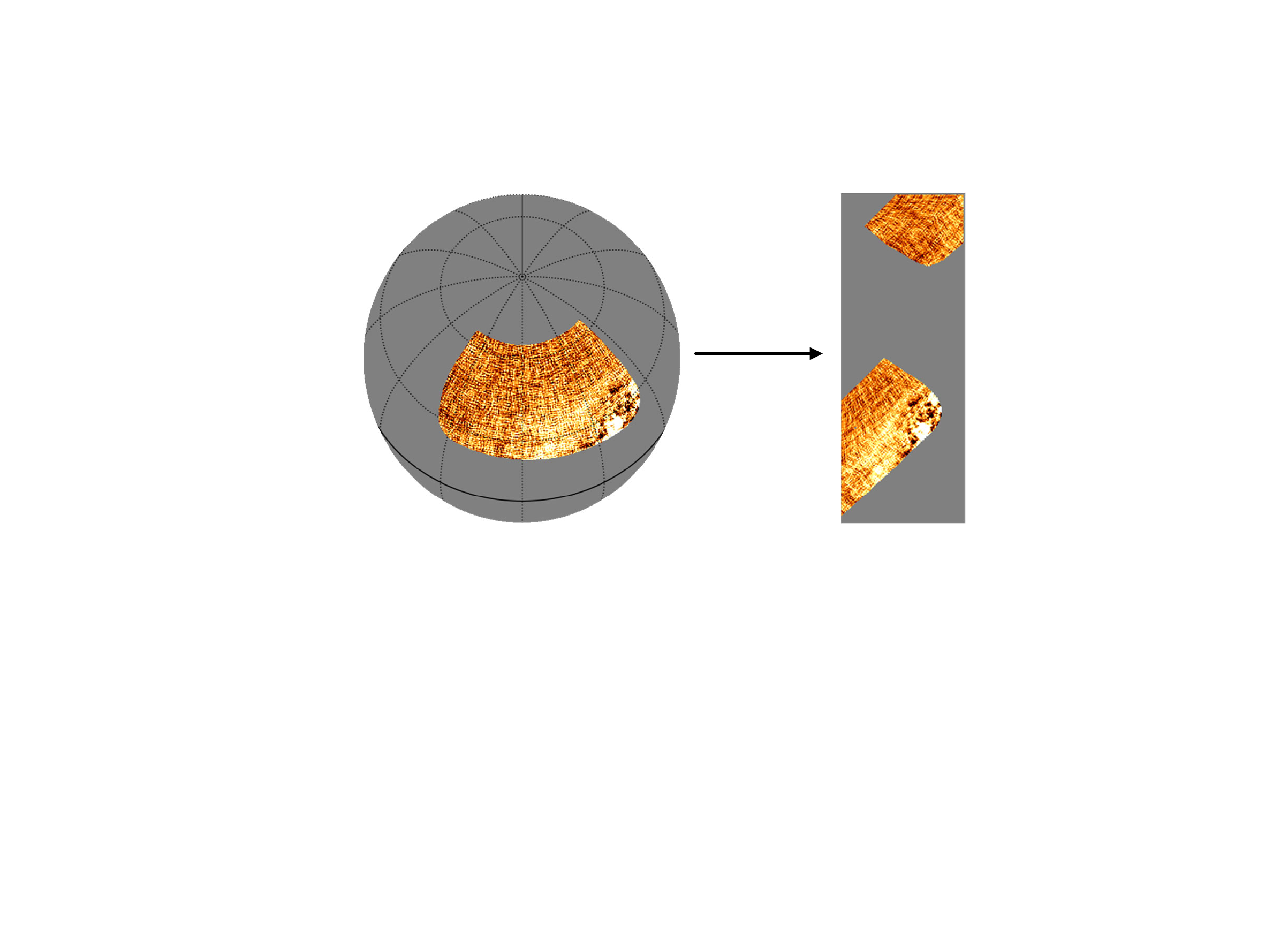}
	\caption{The CNN method requires the transformation of HEALPix spherical sky map into flat map. The regions of the sky that remain unobserved are filled with zeros. }  
	\label{fig:cnn-flat}
\end{figure}

The CNN method is limited to processing flat images, necessitating the conversion of HEALPix spherical sky maps into flat sky maps. As illustrated in Figure \ref{fig:cnn-flat}, the observed sky area is artificially divided into two independent regions within the flat image, introducing non-natural boundaries. In our implementation, we convert the spherical sky maps from the dataset into flat sky maps and use them to train the \texttt{CMBFSCNN} network model. The output maps from the \texttt{CMBFSCNN} network consist of beam-convolved CMB maps combined with noise at 95 GHz.

\begin{figure*}
	\centering
	\subfigure[Recovery of CMB Q map]{
		\includegraphics[width=7cm]{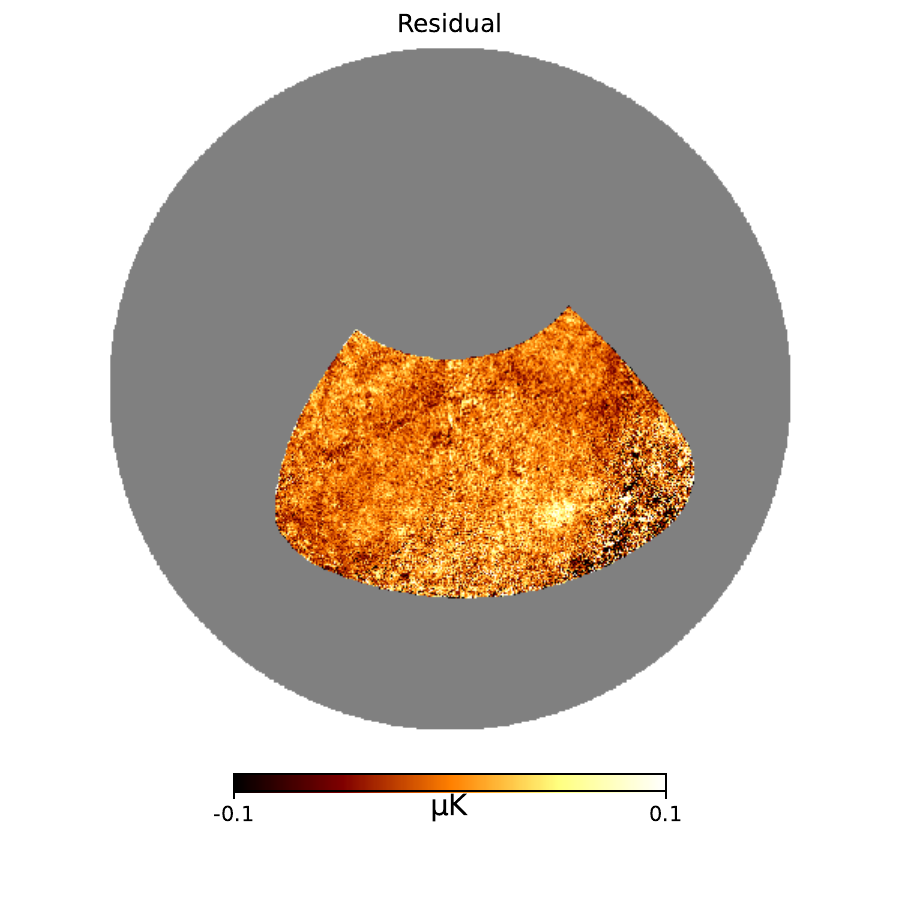}
		%\caption{fig1}
	}
	\vspace{-1mm}
	\hspace{1mm}
	%\quad
	\subfigure[Recovery of CMB U map]{
		\includegraphics[width=7cm]{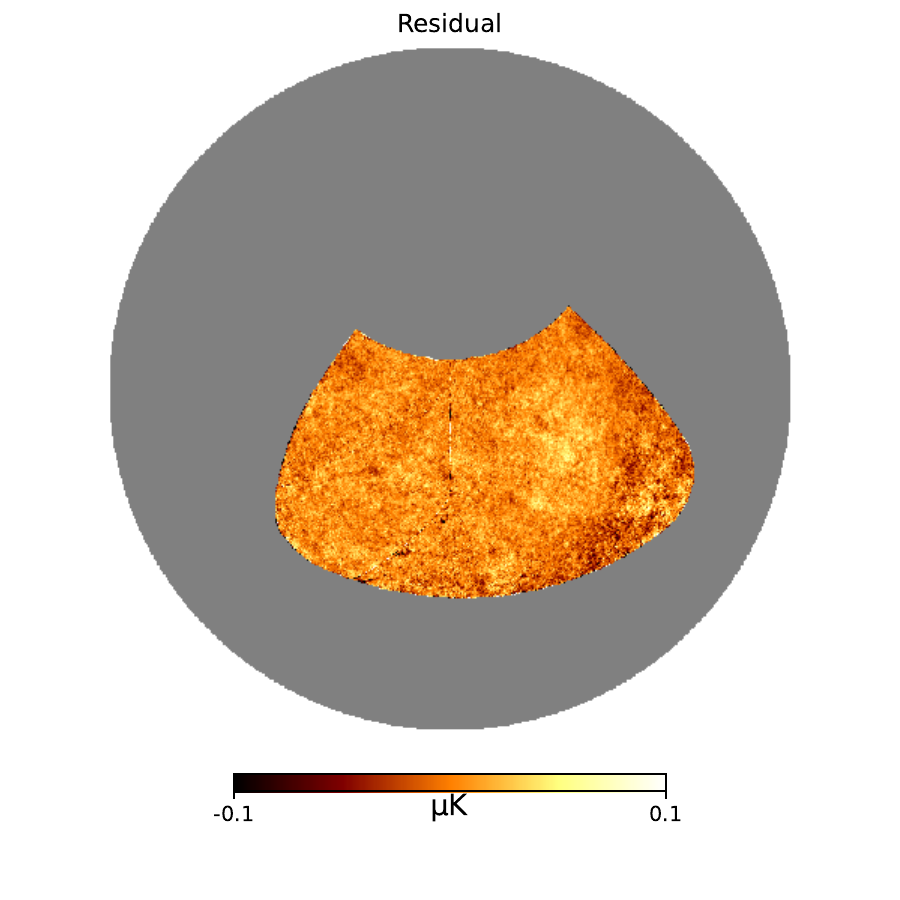}
	}
	\caption{The results of foreground removal using \texttt{CMBFSCNN} method at map level. The residual maps indicate the differences between the recovered noisy CMB map and the target map.}
	\label{fig:recovered-maps-cnn}
\end{figure*}

The results of foreground removal are illustrated in Figure \ref{fig:recovered-maps-cnn}. For clarity, we only present the residual maps of the recovered CMB Q and U, which help highlight the differences between the two methods. Compared to Figure \ref{fig:recovered-maps}, the residual maps show significantly more pronounced artifacts, indicating that the \texttt{CMBFSCNN} method is less effective than the \texttt{TCMB} method in recovering the noisy CMB Q and U maps.

To facilitate a quantitative comparison, we calculated the MAD values between the reconstructed CMB maps and the target sky maps. For the test set, the MAD values are $\sigma_{\rm MAD}^Q = 0.027\pm0.0034$ $\mu$K for recovery of CMB Q map and $\sigma_{\rm MAD}^U = 0.019\pm0.0032$ $\mu$K for recovery of CMB U map. These MAD values are approximately three times larger than those obtained using the \texttt{TCMB} method, and they demonstrate that, at the map level, the \texttt{TCMB} method is more efficient than the \texttt{CMBFSCNN} method for foreground removal.

Furthermore, it is noteworthy that the residual map in Figure \ref{fig:recovered-maps-cnn} exhibits two prominent boundary marks, which arise from the artificially introduced boundaries during the transformation of HEALPix spherical sky maps into flat maps. These additional boundaries disrupt the spatial continuity of the signal, leading to contamination at the discontinuous boundaries during the foreground removal process and resulting in edge effects. This highlights a significant drawback of using the CNN method for processing HEALPix spherical sky maps.
In contrast, the Transformer-based \texttt{TCMB} method can directly process HEALPix spherical sky maps, effectively mitigating edge effects. As a result, the \texttt{TCMB} method demonstrates a clear advantage over the CNN method when handling HEALPix spherical sky maps.

\begin{figure*}
	\centering
	\includegraphics[width=1\hsize]{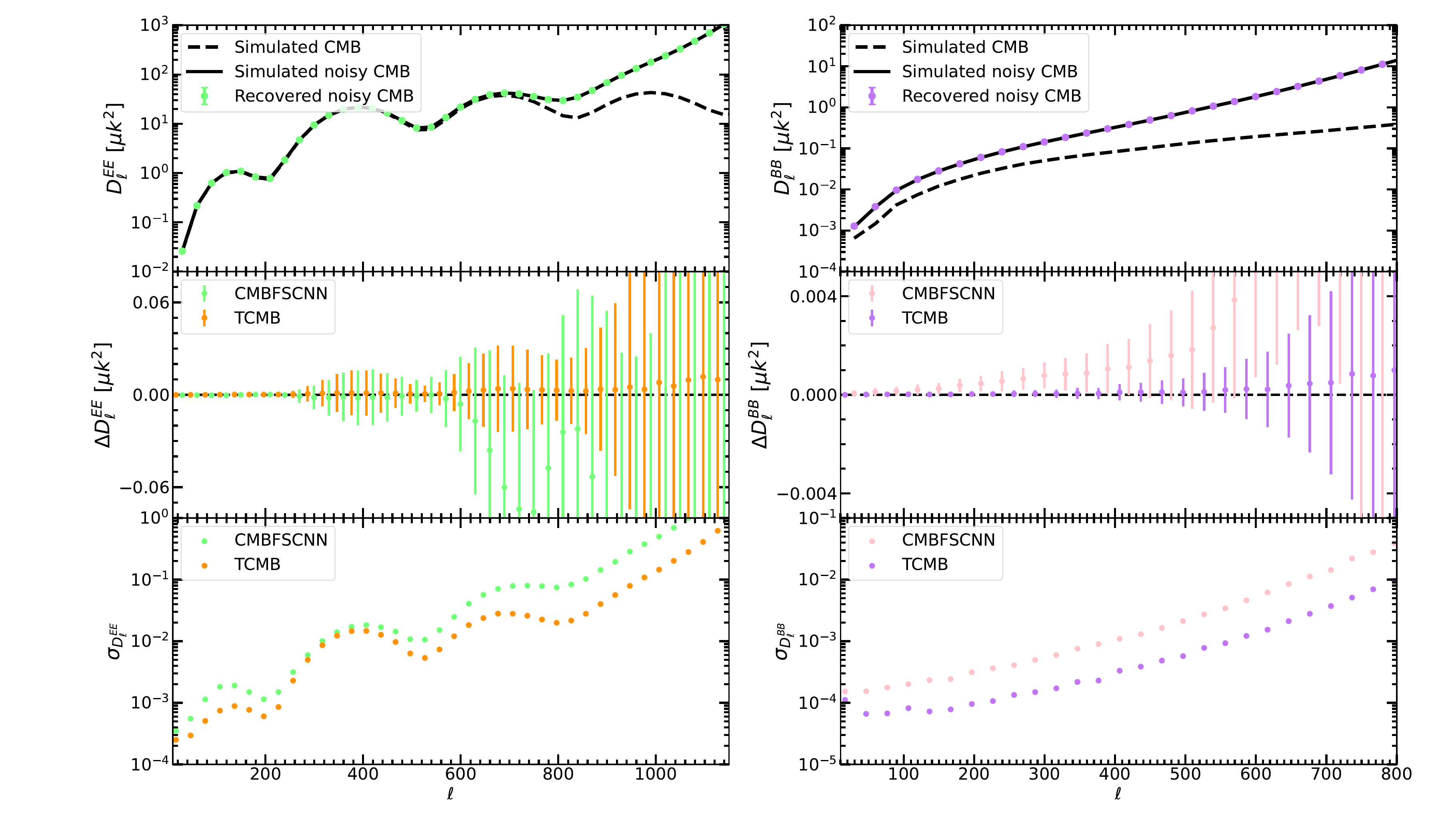}
	\caption{Similar to Figure \ref{fig:noisy_ps}, this Figure presents the results of foreground removal based on the \texttt{CMBFSCNN} method. To facilitate comparison with the \texttt{TCMB} method, we have also included the results obtained using the \texttt{TCMB} method. $\sigma_{D_{\ell}}$ represents the magnitude of the error bars.}  
	\label{fig:noisy-ps-cnn}
\end{figure*}

The power spectra of the noisy CMB maps reconstructed using the \texttt{CMBFSCNN} network are presented in Figure \ref{fig:recovered-maps-cnn}.  
%To facilitate comparison with the \texttt{TCMB} method, the results obtained using the \texttt{TCMB} method are also included in this Figure. 
For comparison, the results obtained using the \texttt{TCMB} method are also included in the figure.
%We can see that the noisy CMB EE and BB power spectra reconstructed by the \texttt{CMBFSCNN} network are in substantial agreement with the target spectra; however, there are minor bises, particularly at smaller scales where these bises are more pronounced.
We observe that the noisy CMB EE and BB power spectra reconstructed by the \texttt{CMBFSCNN} network are in substantial agreement with the target spectra; however, minor biases are present, particularly at smaller scales where these biases are more pronounced.
%It is important to note that these discrepancies remain relatively small in comparison to the noisy CMB power spectra. 
It is important to note that these discrepancies remain relatively small compared to the noisy CMB power spectra.
%In contrast, the CMB power spectrum obtained using the \texttt{TCMB} method exhibits negligible bias. Furthermore, the power spectrum errors recovered by the \texttt{TCMB} method are significantly smaller than those obtained using the \texttt{CMBFSCNN} method across all angular scales. 
In contrast, the CMB power spectrum obtained using the \texttt{TCMB} method exhibits negligible bias. 
%Therefore, from the perspective of the power spectrum, the \texttt{TCMB} method demonstrates a significant advantage over the \texttt{CMBFSCNN} method in the reconstruction of noisy CMB maps.
Furthermore, the uncertainties in the power spectrum recovered by the \texttt{TCMB} method are significantly smaller than those obtained using the \texttt{CMBFSCNN} method across all angular scales. Therefore, from the perspective of the power spectrum, the \texttt{TCMB} method demonstrates a significant advantage over the \texttt{CMBFSCNN} method in the reconstruction of noisy CMB maps.

\begin{figure*}
	\centering
	\includegraphics[width=1\hsize]{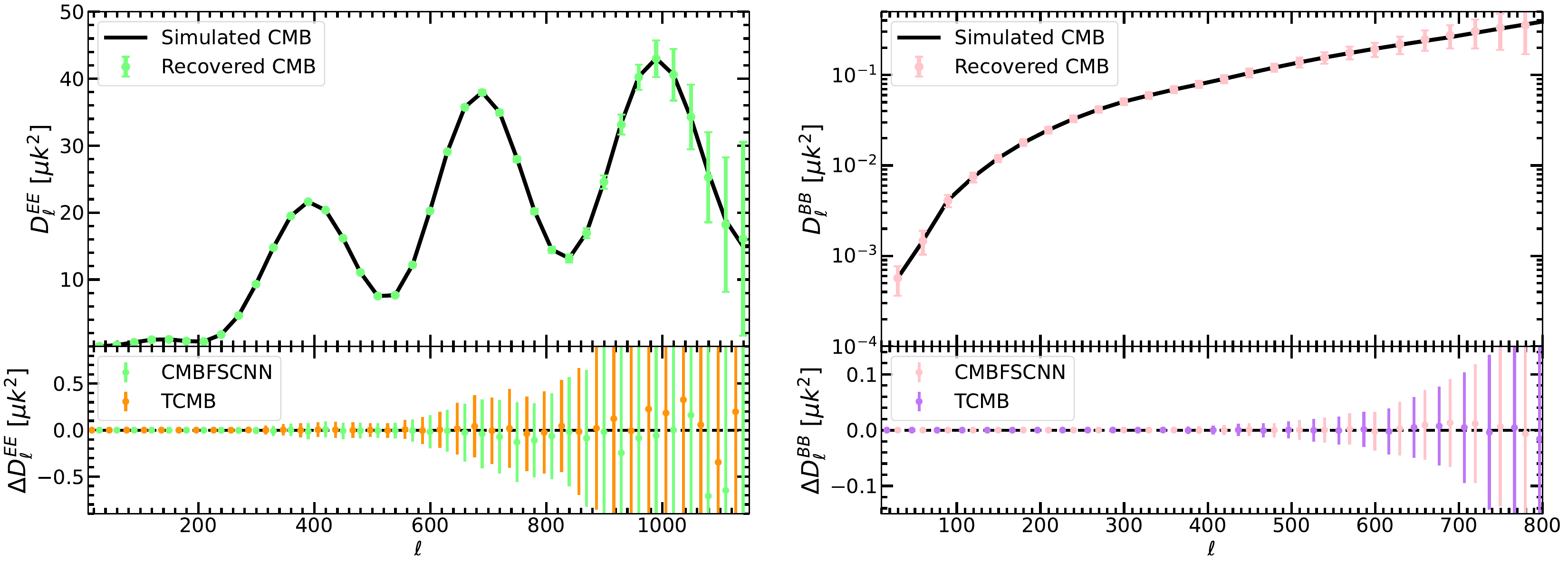}
	\caption{Similar to Figure \ref{fig:recov_ps}, this Figure presents the results of foreground removal based on the \texttt{CMBFSCNN} method. To facilitate comparison with the \texttt{TCMB} method, we have also included the results obtained using the \texttt{TCMB} method.}  
	\label{fig:recov-ps-cnn}
\end{figure*}
\begin{figure*}
	\centering
	\subfigure[Recovery of CMB Q map]{
		\includegraphics[width=8cm]{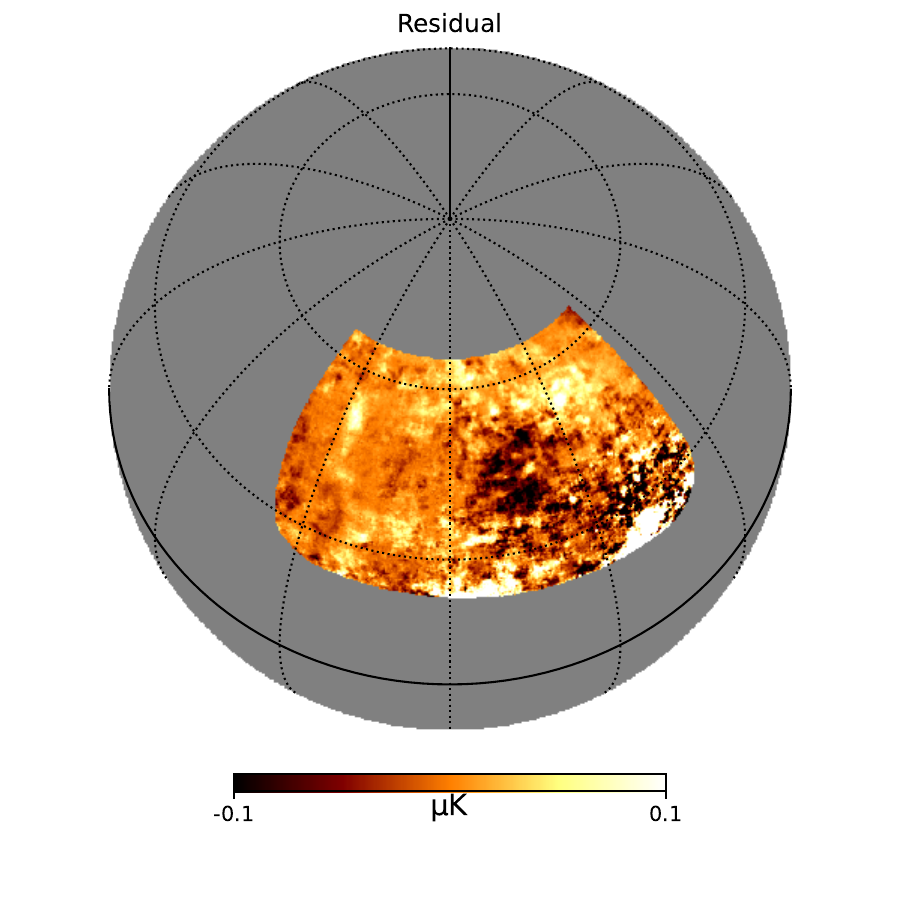}
		%\caption{fig1}
	}
	\vspace{-1mm}
	\hspace{1mm}
	%\quad
	\subfigure[Recovery of CMB U map]{
		\includegraphics[width=8cm]{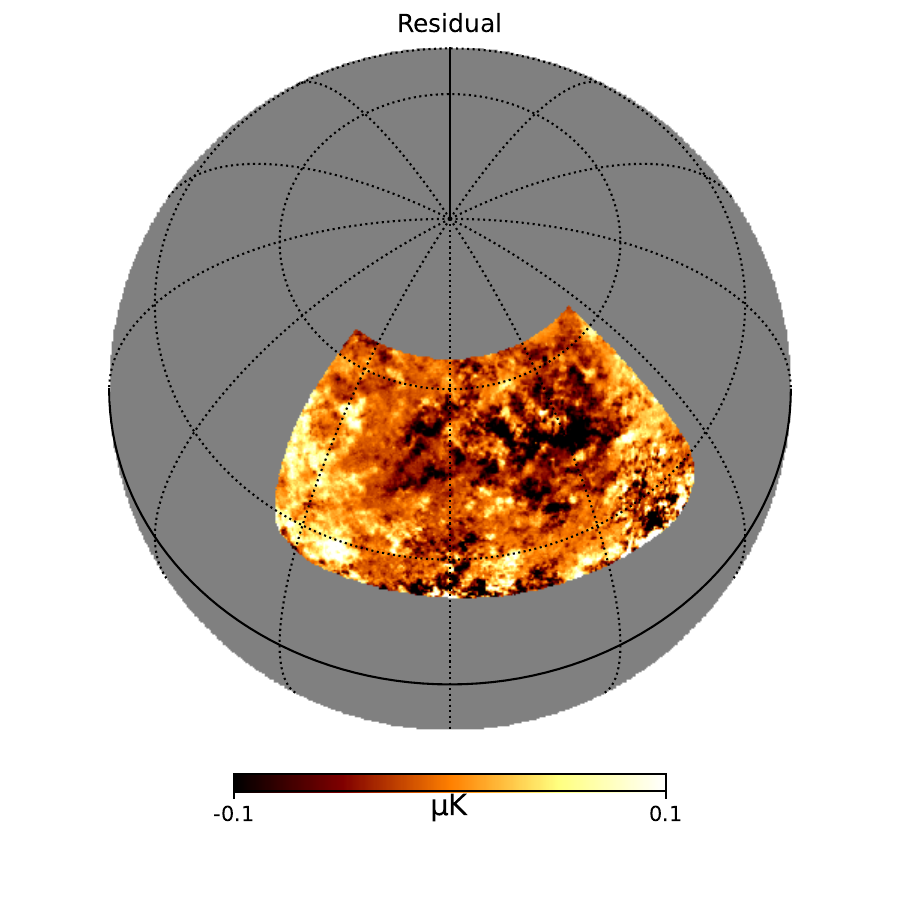}
	}
	\caption{Map-level foreground removal results obtained using the \texttt{TCMB} and \texttt{CMBFSCNN} methods for \texttt{Test Set 1}, where the foreground model in the test dataset exhibits a deliberate mismatch with the training dataset's foreground model}
	\label{fig:recovered-maps-s2d4}
\end{figure*}
\begin{figure*}
	\centering
	\includegraphics[width=1\hsize]{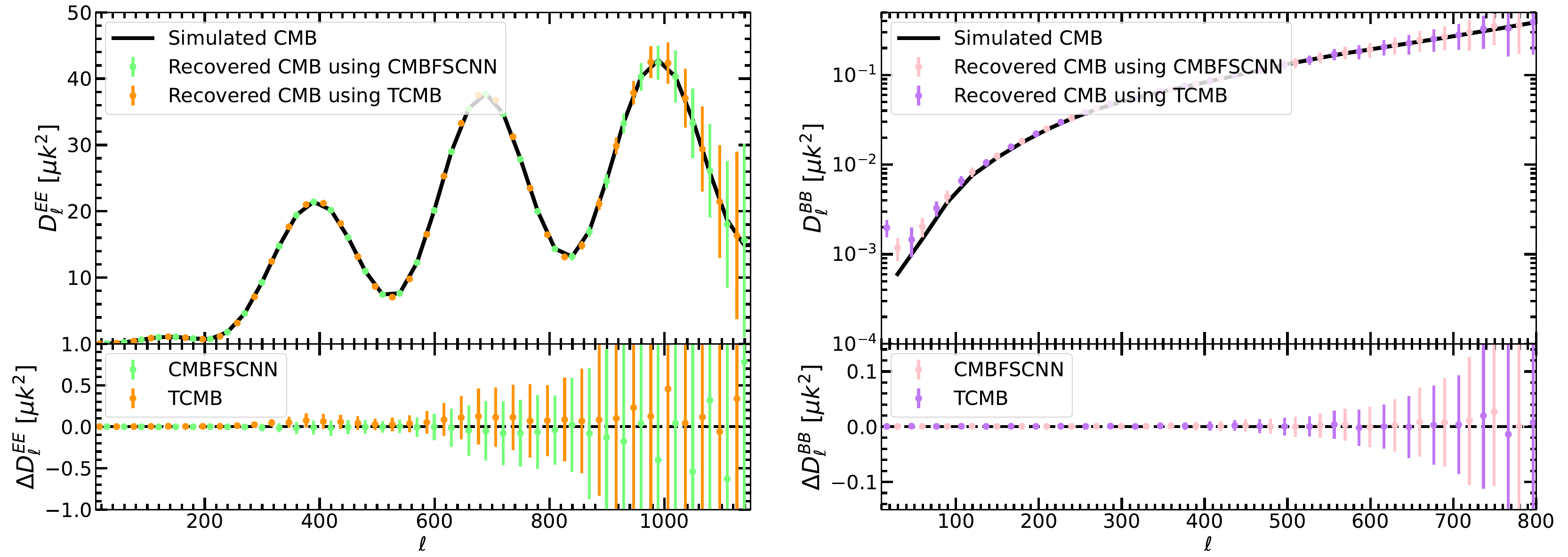}
	\caption{Similar to Figure \ref{fig:recov-ps-cnn}, this Figure presents the recovered power spectra obtained using the \texttt{TCMB} and \texttt{CMBFSCNN} methods for \texttt{Test Set 1}. }  
	\label{fig:noisy-ps-cnn-s2d4}
\end{figure*}

\begin{figure*}
	\centering
	\includegraphics[width=1\hsize]{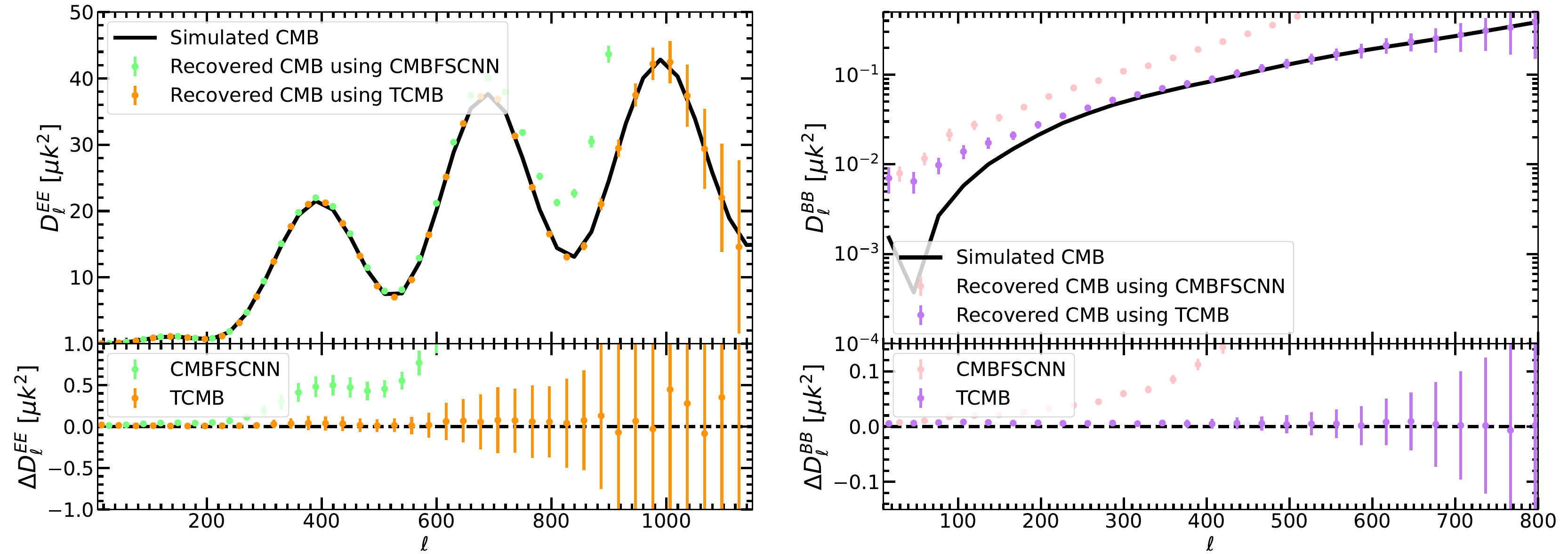}
	\caption{Similar to Figure \ref{fig:recov-ps-cnn},  this Figure presents the recovered power spectra obtained using the \texttt{TCMB} and \texttt{CMBFSCNN} methods for \texttt{Test Set 2}.}  
	\label{fig:denoise-EEBB-cnn-testset-s2d10}
\end{figure*}

\begin{figure*}
	\centering
	\includegraphics[width=1\hsize]{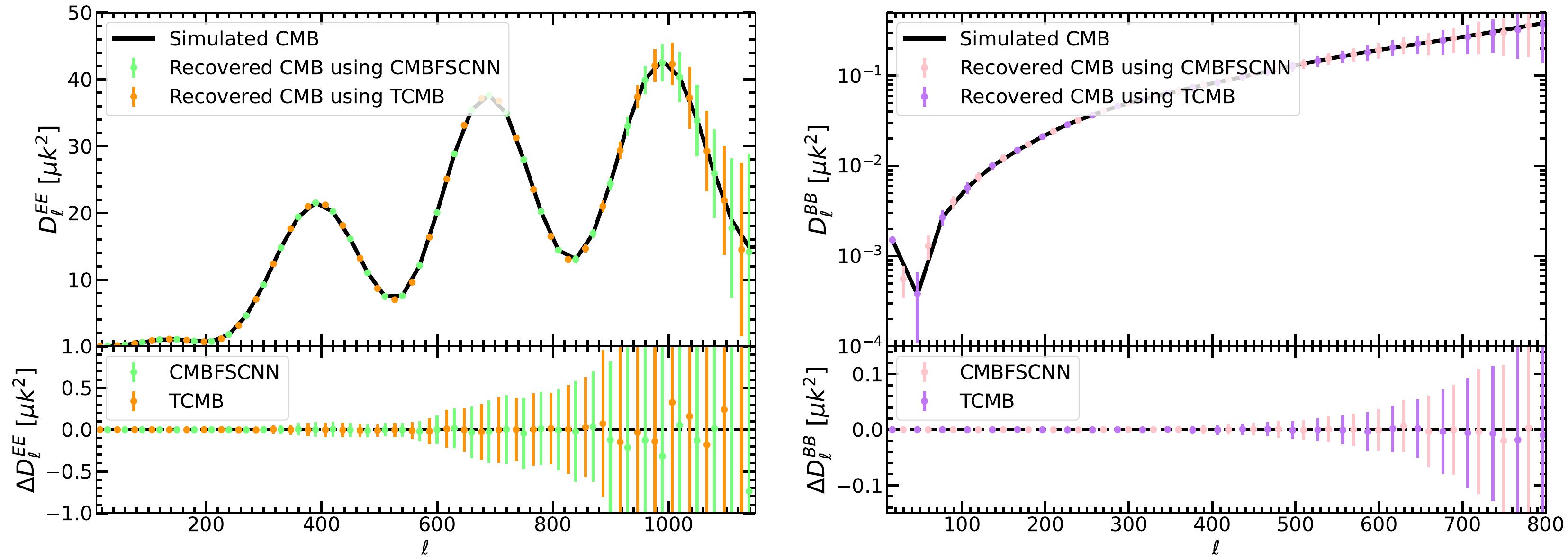}
	\caption{ We augmented the training set by incorporating the d10 thermal dust emission model and retrained the network using this augmented dataset. We applied this trained network to perform foreground removal on \texttt{Test Set 2}.  This Figure presents the recovered power spectra  obtained through this procedure.}  
	\label{fig:denoise-EEBB-cnn-s2d10}
\end{figure*}

We employ cross-correlation analysis of two half-mission maps to eliminate noise biases in CMB power spectra. The results are presented in Figure \ref{fig:recov-ps-cnn}. We can see that the CMB EE and BB power spectra obtained using the \texttt{CMBFSCNN} method are consistent with the true values. In this analysis, no significant advantage of the \texttt{TCMB} method over the \texttt{CMBFSCNN} method in reconstructing the CMB power spectra was identified. This is primarily attributed to the fact that the main contribution to the errors shown in the Figure \ref{fig:recov-ps-cnn} originates from instrument noise. While the \texttt{TCMB} method demonstrates superior performance in reconstructing noisy CMB Q/U maps compared to the \texttt{CMBFSCNN} method, this advantage becomes significantly obscured after the removal of noise biases due to the uncertainties introduced by noise. Consequently, the noise level plays a dominant role in the uncertainty associated with the recovery of the CMB polarization signals. Additionally, during the process of debiasing the power spectrum, the network model does not directly participate, as it primarily processes the sky map data. A more equitable comparison should be based on recovered maps and noisy power spectra as shown Figures \ref{fig:noisy-ps-cnn}. Therefore, the conclusion that the \texttt{TCMB} method outperforms the \texttt{CMBFSCNN} method in removing CMB foregrounds remains valid.

\subsection{Dependency of foreground models}

Deep learning methods inherently depend on the physical models underlying the training data. In this section, we investigate the dependency of the network model on the foreground model. For the training set in this work, we simulate synchrotron emission using the s1 model from the \texttt{PySM} package, while thermal dust emission is modeled via the d1 model. After the network has completed its training, we generate two independent test datasets to evaluate performance under different foreground modeling assumptions, designated as \texttt{Test Set 1} and \texttt{Test Set 2}.  For \texttt{Test Set 1}, we implement the s2 model for synchrotron emission and the d4 model for thermal dust emission. The s2 model introduces a latitude-dependent spectral index variation. Furthermore,  the d1 template employs a single-component modified blackbody spectrum, while the d4 model incorporates a two-component thermal dust parameterization.  \texttt{Test Set 2} utilizes the s2 model for synchrotron emission and adopts the d10 model for thermal dust modeling. The d10 model derives from GNILC needlet-based analysis of Planck data, offering improvements over the Commander-based d1 template: (1) reduced contamination from cosmic infrared background (CIB) and unresolved point sources, and (2) enhanced small-scale structural information preservation.  The two test sets each comprise 2,000 independently generated sky maps, where the variation patterns of cosmological parameters and foreground model parameters remain consistent with the training set, all employing independent sampling.

Following the completion of neural network training on the training dataset, we quantitatively assess the model's dependence on foreground modeling by feeding both \texttt{Test Set 1} and \texttt{Test Set 2} into the trained network architecture. Figure \ref{fig:recovered-maps-s2d4} presents the residual maps between the reconstructed CMB maps obtained through the \texttt{TCMB} methodology and the target maps for \texttt{Test Set 1}. Comparative analysis with the results shown in Figure \ref{fig:recovered-maps} reveals enhanced structural features in the residual maps, suggesting that modifications to the foreground emission model in the test dataset have compromised the network's foreground removal efficacy. This manifests as residual foreground contamination that persists in the output maps. For quantitative evaluation, we computed the MAD metric. The \texttt{TCMB} reconstruction yields $\sigma_{\rm MAD}^Q = 0.0428\pm0.0048$ $\mu$K for recovery of CMB Q map and $\sigma_{\rm MAD}^U = 0.0381\pm0.004$ $\mu$K for recovery of CMB U map in \texttt{Test Set 1}. These values exhibit a statistically slightly  increase compared to the reference results in Figure \ref{fig:recovered-maps}. To benchmark the foreground model dependence, we performed parallel analysis using the \texttt{CMBFSCNN} algorithm. The corresponding MAD values for \texttt{Test Set 1} are $\sigma_{\rm MAD}^Q = 0.050\pm0.007$ $\mu$K for recovery of CMB Q map and $\sigma_{\rm MAD}^U = 0.045\pm0.006$ $\mu$K for recovery of CMB U map.  These values exhibit a statistically slightly  increase compared to the reference results in Figure \ref{fig:recovered-maps-cnn}.

For \texttt{Test Set 2}, we present only the MAD quantification results to evaluate the foreground removal performance at the map level. The \texttt{TCMB} reconstruction yields $\sigma_{\rm MAD}^Q = 0.272\pm0.0355$ $\mu$K for recovery of CMB Q map and $\sigma_{\rm MAD}^U = 0.196\pm0.0235$ $\mu$K for recovery of CMB U map in \texttt{Test Set 2}. The \texttt{CMBFSCNN} reconstruction yields $\sigma_{\rm MAD}^Q = 0.272\pm0.0355$ $\mu$K for recovery of CMB Q map and $\sigma_{\rm MAD}^U = 0.196\pm0.0235$ $\mu$K for recovery of CMB U map in \texttt{Test Set 2}. The analysis reveals significant degradation in foreground removal performance for both \texttt{TCMB} and \texttt{CMBFSCNN} methods when applied to \texttt{Test Set 2} at the map level. The MAD values exhibit an order-of-magnitude increase compared to the results presented in Figure \ref{fig:recovered-maps} and Figure \ref{fig:recovered-maps-cnn}. These findings conclusively indicate that the trained neural network models fail to effectively process \texttt{Test Set 2}, fundamentally because the foreground model variations implemented in Test Set 2 exceed the generalization bounds of the trained networks.

We further present the foreground model dependence of the trained neural network through angular power spectrum analysis. Specifically, we first perform foreground removal on the data of \texttt{Test Set 1} and \texttt{Test Set 2} using the trained network model, then calculate the angular power spectrum of the clean maps. By comparatively analyzing these power spectrum characteristics, we can statistically evaluate the dependence of the network model to different foreground models. Figure \ref{fig:noisy-ps-cnn-s2d4} presents a comparative analysis of the recovered CMB power spectra obtained through the \texttt{TCMB} and \texttt{CMBFSCNN} methodologies for \texttt{Test Set 1}. Our results demonstrate that while both methods successfully reconstruct the essential features of the CMB power spectra, the BB power spectrum exhibits slight deviations at large angular scales ($\ell<70$), which we attribute to residual foreground contamination. In contrast, the EE power spectrum reconstruction remains remarkably robust. This differential sensitivity between polarization modes highlights the greater vulnerability of BB-mode measurements to foreground model, consistent with theoretical expectations given the intrinsically weaker amplitude of the  BB-mode signal compared to EE modes.

Figure \ref{fig:denoise-EEBB-cnn-testset-s2d10} presents a comparison of the reconstructed CMB power spectra using both the  \texttt{TCMB} approach and  \texttt{CMBFSCNN} method for \texttt{Test Set 2}.  The results reveal significant limitations in the \texttt{CMBFSCNN} method: its reconstructed EE-mode power spectrum shows substantial deviations from the true spectrum at angular scales $\ell>600$, while the reconstructed BB-mode power spectrum exhibits significantly biases across all angular scales. This indicates that the trained \texttt{CMBFSCNN} model fails to achieve precise foreground removal when processing \texttt{Test Set 2}.
In contrast, the \texttt{TCMB} method demonstrates superior robustness in EE-mode power spectrum reconstruction, showing relatively less susceptibility to variations in foreground modeling. However, it still exhibits  deficiencies in BB-mode reconstruction, particularly showing slight deviations at angular scales ($\ell<300$). 

\begin{figure*}
	\centering
	\includegraphics[width=1\hsize]{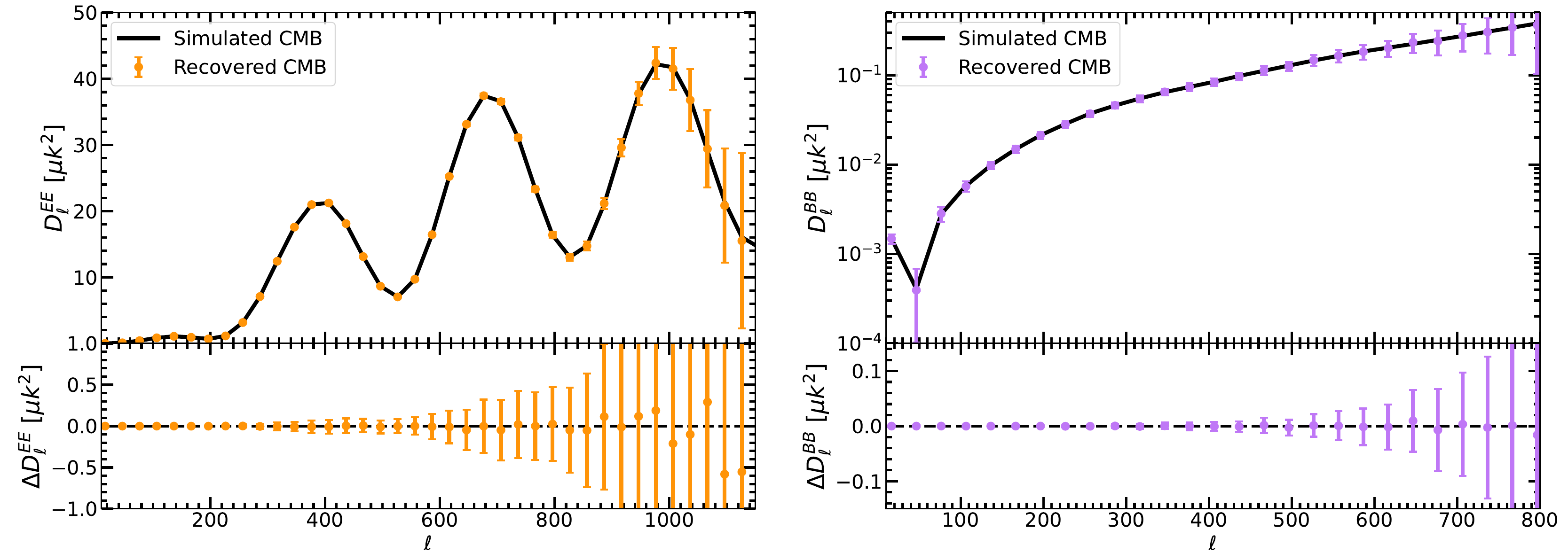}
	\caption{Similar to Figure \ref{fig:recov_ps},  however, the noise levels in the testing dataset have been altered in this Figure.}  
	\label{fig:noisy-ps-cnn-noisechange}
\end{figure*}

A comparative analysis of the \texttt{TCMB} and \texttt{CMBFSCNN} methods reveals that \texttt{TCMB} exhibits superior robustness in foreground removal performance. This enhanced stability primarily stems from two key factors: (1) \texttt{TCMB}'s significantly larger model parameter space, and (2) its utilization of more extensive training datasets.  These characteristics collectively enable \texttt{TCMB} to maintain better performance stability when handling complex foreground variations, demonstrating particular resilience in preserving the spectral characteristics of both EE and BB polarization modes across different angular scales.

In the current study, the simulation of the training dataset employs only a single foreground model configuration, specifically using the s1 model for synchrotron radiation and the d1 model for thermal dust emission. However, actual observational data often exhibit more complex properties, and the singular combination of s1 and d1 models could not fully capture these observed features. Considering that deep learning methods are inherently data-driven modeling processes, their performance largely depends on the physical model of the training dataset. To enhance the generalization capability of the deep learning model when processing real observational data, we propose introducing more diverse foreground model combinations during the construction of the training set. Specifically, we expanded the original training set by 30\% in sample size, with these additional samples adopting different physical model configurations: the s2 model for synchrotron radiation simulation and the d10 model for thermal dust emission simulation. This improvement strategy ensures that the final training set incorporates two distinct foreground model combinations (s1+d1 and s2+d10).

During the model training phase, we optimized the network parameters based on this enhanced training set. After completing the model training, we evaluated the performance of the trained network in foreground removal tasks using an independent test dataset (\texttt{Test Set 2}). At the map level, the reconstruction performance is evaluated using the MAD metric. For \texttt{Test Set 2}, the \texttt{TCMB} method achieves a reconstruction precision  $\sigma_{\rm MAD}^Q = 0.0096\pm0.0038$ $\mu$K for recovery of CMB Q map and $\sigma_{\rm MAD}^U = 0.0087\pm0.0041$ $\mu$K for recovery of CMB U map in \texttt{Test Set 2}. These value are consistent with the results presented in Section \ref{sec:recoveredCMB}. Notably, the \texttt{CMBFSCNN} approach demonstrates identical reconstruction accuracy, yielding  $\sigma_{\rm MAD}^Q = 0.029\pm0.0041$ $\mu$K for recovery of CMB Q map and $\sigma_{\rm MAD}^U = 0.024\pm0.0039$ $\mu$K for recovery of CMB U map in \texttt{Test Set 2}. The MAD values show agreement with the corresponding results in Sections \ref{sec:CNN}. Figure \ref{fig:denoise-EEBB-cnn-s2d10} presents the reconstructed CMB power spectra obtained using two different methods. The results demonstrate that both the EE and BB polarization power spectra are accurately recovered. These results indicate that when the training dataset incorporates thermal dust emission simulated with the d10 model and synchrotron radiation simulated with the s2 model, the trained neural network effectively removes foreground contamination in \texttt{Test Set 2}. This finding confirms that expanding the range of foreground models in the training dataset significantly enhances the generalization capability and foreground removal performance of the neural network model.

To effectively reduce neural networks' dependence on specific foreground models, we can implement the following improvements at the data level. For instance, we can construct more diverse training datasets by integrating sample data generated from multiple foreground models to enhance data heterogeneity. Additionally, we can employ a weighted linear combination approach to fuse sky maps from different foreground models, thereby generating synthetic foreground maps with broader coverage. These methods will significantly improve the statistical diversity of training samples, thereby mitigating the risk of overfitting to specific foreground features. It is particularly noteworthy that the scientific construction of astrophysical datasets suitable for deep learning remains a critical research challenge, which will be a key focus of future work.

\subsection{Variation in the level of instrument noise}
In actual observations, the level of instrumental noise could fluctuate. In this study, we investigate the impact of variations in instrumental noise levels on the results. After the network model has been trained on the standard training set, we re-simulate the testing dataset. Specifically, for the noise standard deviation maps of AliCPT at 95 GHz and 150 GHz, we multiply the value of each pixel by a random number drawn from a distribution with a mean of 1 and a variance of 0.1. In other words, each pixel in the noise standard deviation map is independently adjusted by 10\%. We input the testing dataset, altered to reflect variations in noise levels, into the network model that has been trained on the standard training dataset, in order to evaluate the impact of noise level fluctuations on the model's performance.

The reconstructed CMB maps are found to be largely consistent with the results presented in Figure \ref{fig:recovered-maps}. Here, we do not display specific results at map level; rather, we provide quantitative measures of the mean absolute deviation (MAD) to evaluate the effectiveness of foreground removal. The MAD value for the reconstructed CMB Q map is observed to be $0.0084\pm0.0038\ \mu$K, while the MAD for the U map is $0.0079\pm0.0035\ \mu$K. These MAD values are consistent with the results presented in Section \ref{sec:recoveredCMB}. After the removal of noise bias, Figure \ref{fig:noisy-ps-cnn-noisechange} presents the reconstructed CMB polarization power spectra. It is evident that the reconstructed CMB polarization power spectrum is consistent with the results shown in Figure \ref{fig:recov_ps}. These results suggest that minor variations in noise levels for each pixel have a negligible impact on the outcomes.

\section{Conclusions}\label{sec:conclusions}
In this study, we propose a deep learning method based on the Transformer model, called \texttt{TCMB}, which is capable of effectively processing HEALPix spherical sky map. This method is applied to the foreground removal of CMB polarization in the ground based CMB polarization observations. The results demonstrate that our method can efficiently remove the CMB foreground, validating its effectiveness.

Based on the simulated data from the AliCPT-1 like experiment, we employed the \texttt{TCMB} method to perform foreground removal of CMB polarization. At the sky map level, the \texttt{TCMB} method effectively removes the foreground while retaining the CMB signal and instrument noise. We computed the cross-correlation between two half-mission maps to eliminate noise bias, and the results demonstrate that the \texttt{TCMB} method can effectively recover the CMB polarization signal, with the recovered EE and BB power spectra consistent with the true values.

We compared the \texttt{TCMB} method with the \texttt{CMBFSCNN} approach previously employed based on the CNN, and the results indicate that \texttt{TCMB} outperforms \texttt{CMBFSCNN} in recovering the CMB maps, while also overcoming the edge effects associated with the inability of CNN methods to directly process Healpix-format spherical sky maps. This demonstrates that the \texttt{TCMB} method has a distinct advantage over CNN approaches when dealing with Healpix-format spherical sky maps.

Additionally, we investigated the model dependency of \texttt{TCMB}, and the results indicate a clear dependence of \texttt{TCMB} on the foreground model. In scenarios where the foreground model undergoes significant changes, the efficiency of foreground removal may decrease; however, the impact of the residual foreground on the results is limited. Finally, we tested the effect of variations in instrument noise levels on the results, and the findings suggest that a 10\% change in instrument noise levels has a negligible impact on the outcomes.

The \texttt{TCMB} method can effectively handle HEALPix spherical sky maps, making it applicable not only to the processing of CMB data but also to other experimental datasets, such as 21 cm signal data and galaxy survey data. 

We thank Yi-Ming Wang, Yong-Ping Li, Hua Zhai, Wen-Zheng Chen for useful discussion. This study is supported by the Postdoctoral Fellowship Program of CPSF under Grant Number GZC20241729, the China Postdoctoral Science Foundation under Grant Number 2024M763267, the National Nature Science Foundation of China under grant Nos. 12403005 12473004, and 12021003, the National Key R\&D Program of China No. 2020YFC2201601 and 2020YFC2201603, and nhe Fundamental Research Funds for the Central Universities. Some of the results in this paper have been derived using the HEALPix (page: \url{https://healpix.sourceforge.io/}). We would like to express our gratitude for the following software packages: \texttt{Healpy}, \texttt{CAMB}, \texttt{NaMaster}, \texttt{LensPyx}, \texttt{PySM}, and \texttt{PySM3}.

%%%%%%%%%%%%%%%%%%%%%%%

\bibliography{ref}{}
\bibliographystyle{aasjournal}

% \bibliography{sample63.bib}{}
% \bibliographystyle{aasjournal.bst}

%\begin{thebibliography}{}
%\end{thebibliography}

%% This command is needed to show the entire author+affiliation list when
%% the collaboration and author truncation commands are used.  It has to
%% go at the end of the manuscript.
%\allauthors

%% Include this line if you are using the \added, \replaced, \deleted
%% commands to see a summary list of all changes at the end of the article.
%\listofchanges

\end{document}